\documentclass{aa}
\usepackage[varg]{txfonts}

\usepackage{natbib}
\bibpunct{(}{)}{;}{a}{}{,} 

\usepackage{graphicx}

\usepackage[]{hyperref}

\begin{document}

   \title{Spatiotemporal evolution of UV pulsations and their connection to 3D magnetic reconnection and particle acceleration}


\author{
    Stefan Purkhart\inst{1},
    Hannah Collier\inst{2,3},
    Laura A. Hayes\inst{4},
    Astrid M. Veronig\inst{1}\fnmsep\inst{5},
    Miho Janvier\inst{6},
    Säm Krucker\inst{2}\fnmsep\inst{7}
    }

\institute{
    Institute of Physics, University of Graz, Universitätsplatz 5, 8010 Graz, Austria\\
    \email{stefan.purkhart@uni-graz.at}
    \and
    University of Applied Sciences and Arts Northwestern Switzerland (FHNW), Bahnhofstrasse 6, 5210 Windisch, Switzerland
    \and
     ETH Z\"{u}rich,
        R\"{a}mistrasse 101, 8092 Z\"{u}rich, Switzerland
    \and
    Astronomy \& Astrophysics Section, School of Cosmic Physics, Dublin Institute for Advanced Studies, DIAS Dunsink Observatory, Dublin, D15 XR2R, Ireland.
    \and
    Kanzelhöhe Observatory for Solar and Environmental Research, University of Graz, Kanzelhöhe 19, 9521 Treffen, Austria
    \and
    European Space Agency, ESTEC, Noordwijk, The Netherlands
    \and
    Space Sciences Laboratory, University of California, 7 Gauss Way, 94720 Berkeley, USA
        }

\date{Received ; accepted}

\abstract
{
Quasi-periodic pulsations (QPPs) are a common feature of impulsive solar flare emissions, yet their driving mechanism(s) remain unresolved. Observational challenges such as image saturation during large flares, low image cadence, and limited spatial resolution often hinder our ability to study the spatiotemporal evolution of QPPs in detail. The M3.7 solar flare of February 24, 2023 produced long-period (3.4~min) QPPs in hard X-ray (HXR) and ultraviolet (UV) emissions and was associated with a large-scale asymmetric filament eruption. 
}
{
This study leverages the unique opportunity presented by the combination of a long pulsation period and the large spatial scale of the eruptive event. Our goal is to pinpoint the location of the observed QPPs within the flare structure, characterize their spatiotemporal evolution, examine their connection to particle acceleration, and understand their relationship to flare ribbon dynamics and the underlying magnetic reconnection geometry.
}
{
The UV emissions of the flare ribbons were analyzed with SDO/AIA 1600 Å base-difference imaging, while the HXR emissions were studied with Solar Orbiter/STIX. The flare region was divided into four subregions focused on the main flare ribbons or filament footpoints in both polarities, from which the integrated light curves were extracted. Time-distance plots were constructed to follow the motion of the UV ribbon kernels and to relate them to the QPPs. A spectral fitting of the HXR observations was performed to characterize the accelerated electron populations and the HXR images were reconstructed to study the X-ray sources.
}
{
We find that the strength and characteristics of the UV pulsations varied significantly across subregions, with the strongest correlation to HXR pulsations and the most complex spatiotemporal evolution occurring in the compact southern flare ribbon. In this location, UV pulsations originated from a sequence of UV kernels at an expanding flare ribbon front, accompanied by secondary motions along the front. In contrast, another expanding ribbon front, likely associated with a different subsystem of the reconnecting flux, did not exhibit pulsations. The UV pulsations spatially coincided with HXR sources, indicating non-thermal electrons as a common driver. The HXR spectrum exhibited a soft-hard-soft evolution, suggesting a modulation of the electron acceleration efficiency on the same timescale as the pulsations. Around the eastern filament anchor region, UV pulsations were produced instead by plasma injections into a separate filament channel, highlighting the presence of distinct emission sources beyond flare kernels.
}
{
Our results suggest that the spatiotemporal evolution of the UV pulsations is largely driven by the propagation of magnetic reconnection, including slipping reconnection, within an asymmetric magnetic geometry. We further hypothesise that slipping reconnection, combined with the structure of the quasi-separatrix layers (QSLs), plays a key role in the generation and propagation of UV kernels. An additional time-varying reconnection process was potentially required to fully explain the observations. Our findings emphasize the importance of spatially resolved QPP studies on the timescales of energy release to disentangle different emission processes and their connection to magnetic reconnection.
}


\titlerunning{HXR pulsations during the February 24, 2023, M3.7 flare}
\authorrunning{Purkhart et al.}

\maketitle

\section{Introduction}\label{sec:introduction}

Solar eruptive events, which include solar flares and coronal mass ejections (CMEs), are driven by the sudden release of stored magnetic energy through a process of magnetic reconnection. These events involve a dynamic restructuring of the solar magnetic field, which result in a significant energy release in the form of radiation, plasma heating, plasma flows, and the acceleration of energetic particles \citep{Priest2002,Fletcher2011,Benz2017}. An important observational signature of the dynamic processes occurring in these events is that the emission associated with solar flares often show time-dependent variations and pulsations, commonly referred to as quasi-periodic pulsations \citep[QPPs; see][]{Nakariakov2009,Kupriyanova2020}. These pulsations have been reported across a wide range of wavelengths \citep{Hayes2019, Hayes2020,french_2024} and while the term QPP is quite generic and used to describe a variety of time-dependent phenomena, in this study, we focus on impulsive-phase QPPs in flare emissions associated with particle acceleration. The QPPs we consider here are characterized by quasi-periodic enhancements in emission intensity and they are typically seen in hard X-rays, UV, and radio emissions \citep{ sijie_2020, Clarke2021, Collier2024}. They are believed to reflect intrinsic processes within the flare, such as periodic reconnection or bursty particle injections. However, the exact mechanisms driving QPPs remain debated, with proposed models including magnetohydrodynamic (MHD) oscillations, periodic reconnection, and bursty particle injections \citep{McLaughlin2018,Zimovets2021}. Understanding their spatial and temporal evolution is crucial for linking the dynamics of magnetic reconnection to particle acceleration.

Key insights on solar flares can be gained by the observations of flare ribbons, which are elongated brightenings observed in the chromosphere that trace the footpoints of magnetic field lines undergoing reconnection \citep{Fletcher2011}. They are often accompanied by compact, transient brightenings known as flare kernels \citep[e.g.,][]{Kaempfer1983,Asai2004}, which  are a result of the rapid heating of the chromospheric plasma by the collisions of non-thermal flare electrons with the ambient electron population. Therefore, they often coincide with the locations of non-thermal HXR bremsstrahlung emission that is due to the deceleration of the same flare-accelerated electrons in the field of the ambient ions \citep{Brown1971,Holman2011}.

In the standard 2D flare model framework, also known as the CSHKP \citep{Carmichael1964,Sturrock1966,Hirayama1974,Kopp1976} model, flare ribbons are thought to map the progression of magnetic reconnection along a current sheet. However, observations reveal that ribbons and kernels often display complex, time-dependent behavior, which cannot be explained  by the 2D framework in a straightforward way. The inherent 3D complex magnetic structures of flares must be taken into account when trying to understand solar flare ribbons and time-dependent flare emission \citep{Janvier2017}. Processes such as slipping reconnection, in which magnetic field lines appear to slip through the plasma as they change connectivity, have been proposed to explain these dynamics \citep{Dudik2014,Janvier2013}. Slipping reconnection highlights the importance of 3D magnetic structures in modulating the observed flare emissions, linking the evolving ribbons and kernels to particle acceleration processes.

In this context, an important but unresolved question is whether the time-dependent behavior of QPPs in solar flare emission is inherently linked to the spatial and temporal evolution of flare ribbons and kernels. If QPPs are driven by bursty or time-dependent reconnection processes occurring in a 3D magnetic topology, they may reflect fundamental properties of how energy is released and particles are accelerated during solar flares. However, the exact nature of this connection remains unclear. Understanding how QPPs relate to the evolution of flare ribbons and kernels is crucial for linking the dynamics of magnetic reconnection to large scale energy release and particle acceleration. Addressing these open questions requires spatially and temporally resolved observations that can disentangle competing mechanisms. 

One of the challenges in studying QPPs is the difficulty involved in the spatial analysis of the associated temporally observed pulsations. For example, during large flares, emissions in EUV and UV wavebands often saturate; typically, the short periods (i.e., tens of seconds) of many QPPs are difficult to resolve with available EUV/UV cadences \citep{Inglis2023}. As a result, the relationship between UV/EUV brightenings and HXR pulsations, which are critical for understanding the link between the modulation of QPPs and particle acceleration and chromospheric response, have been poorly constrained. The February~24, 2023, eruptive M3.7~flare presents a unique opportunity to address these challenges. The flare was associated with a large-scale asymmetric filament eruption, resulting in a total extent of approximately 50 degrees in longitude. The SOHO/LASCO coronagraph shows that the filament eruption resulted in a fast halo CME with a mean speed of 1300~km/s\footnote{LASCO CME catalog: https://cdaw.gsfc.nasa.gov/}. The event exhibited long-period pulsations (approximately 3.4~minutes) in both HXR and UV emissions, which were observed by Solar Orbiter/STIX and SDO/AIA. These long-period pulsations, combined with the large spatial scale of the flare, enabled a detailed analysis of the temporal and spatial evolution of flare emissions without the usual complications of saturation or cadence constraints. We focus on an analysis to identify how QPPs observed with STIX and AIA are spatially distributed across different regions of chromospheric signatures of the solar eruptive event, including the flare ribbons and evolving filament channels.

The paper is organized as follows:
data acquisition, preprocessing, and analysis are described in Section \ref{sec:methods}. 
In Section \ref{sec:results}, we present our results with the period and spectral analysis of the HXR pulsations (Section \ref{sec:results_STIX_pulsations}) and the introduction of the main event features (Section \ref{sec:results_SDO_GONG_overview}).
The UV pulsation results in Section \ref{sec:results_AIA1600} introduce the four subregions of interest and show their different UV emission (Section \ref{sec:results_AIA1600_light_curves}), the overall spatial evolution of the UV emission in each subregion and the derived parameters (Section \ref{sec:results_AIA1600_masks_and_parameters}), and the detailed spatial and temporal evolution of the UV flare kernels (Section \ref{sec:results_AIA1600_time_dist}), 
The spatial relation between HXR sources and UV kernels is presented in Section \ref{sec:results_STIX_imaging}.
In our discussion in Section \ref{sec:discussion}, we highlight the implications of key observations (Section \ref{sec:discussion_key_findings}), relate them to the propagation of magnetic reconnection (Section \ref{sec:discussion_3D_flare}), identify possible pulsation mechanisms (Section \ref{sec:discussion_pulsation_models}), and discuss their implications (Section \ref{sec:discussion_implications}).
A conclusion is given in Section \ref{sec:conclusion}.

\section{Data and methods}\label{sec:methods}

\subsection{Solar Orbiter/STIX}

The STIX instrument \citep{Krucker2020_STIX} on board Solar Orbiter is a HXR spectrometer and imager observing with a full-Sun field of view in the 4–150 keV range. This range covers bremsstrahlung emission from high-energy non-thermal electrons as well as thermal emission from plasma exceeding 10 MK. For this event, STIX provides key diagnostics on HXR pulsations and footpoint locations, allowing us to analyze the modulation of particle acceleration and the spatial distribution of non-thermal sources.

We downloaded the STIX data for this event from the STIX website\footnote{\url{https://datacenter.stix.i4ds.net}} which is a user-friendly way of interacting with the powerful STIX data platform \citep{Xiao2023}. A background measurement\footnote{UID 2302275471} was subtracted from the data to isolate the counts caused by emission from the event under study. A correction for the difference in light travel time between the Sun and Solar Orbiter and that between the Sun and Earth of 108.6~s was applied to all STIX data. Any time given in this paper is therefore for an Earth based observer, making it easier to compare observations between SDO/AIA and Solar Orbiter/STIX.

The spectral fitting was performed using OSPEX\footnote{\url{http://hesperia.gsfc.nasa.gov/ssw/packages/spex/doc/}} available within the SSWIDL environment. The thermal, lower-energy part of the spectrum was fitted using the isothermal "vth" model and the non-thermal higher energy component was fitted using the thick-target "thick2" model. In addition, an albedo component was included to account for the reflection of some X-ray emission into the observer direction by Compton backscattering in the photosphere \citep{Tomblin1972,Santangelo1973}.

STIX is an indirect imager \citep{Massa2023}, meaning that images have to be reconstructed using one of several developed algorithms. For this study, we used the mem\_ge method \citep{Massa2020} available within the imaging pipeline of the STIX SSWIDL package (\textit{stx\_mem\_ge}). Image coordinates get corrected based on the STIX aspect system \citep{Warmuth2020}, which measures the orientation of STIX relative to the Sun. However, this system is often not fully sufficient, especially when STIX is farther away from the Sun. To achieve proper alignment, we apply an additional manual shift ($\Delta x = -12$~arcsec, $\Delta y = -6$~arcsec) to the images by aligning the STIX non-thermal sources with the AIA 1600~Å ribbons (reprojected to a Solar Orbiter view). This can be done since HXR and UV footpoints are usually the same and is the standard procedure performed in multipoint flare studies involving STIX \citep[e.g.,][]{Collier2024,Purkhart2023}.

Because Solar Orbiter and SDO observed the event from different vantage points, the data must be reprojected into a common reference frame for accurate comparison. For this purpose, we used the reprojection capabilities available in version 5.0.0 \citep{SunPy_5.0.0} of the SunPy open source software package \citep{sunpy_community2020}. While a reprojection for Solar Orbiter/STIX footpoint sources to the SDO/AIA frame is straightforward, coronal source reprojection is more challenging due to insufficient constraints on source height. Instead, we constructed a line-of-sight (LOS) from STIX through the brightest pixel of a given coronal source and then showed this LOS in the AIA images. These lines end where they intersect the photosphere and give a good sense of the relevant EUV features.

\subsection{SDO/AIA and HMI}\label{sec:methods_AIA_HMI}

We used data from the Atmospheric Imaging Assembly \citep[AIA,][]{Lemen2012_AIA} on board the Solar Dynamics Observatory \citep[SDO,][]{Pesnell2012_SDO}, which captures full-disk solar images with a spatial resolution of 1.5~arcsec (a pixel scale of 0.6~arcsec) and a cadence of 12~s for EUV channels and 24~s for UV channels, such as AIA 1600~Å.

The images were downloaded as tracked submaps via the JSOC cutout feature\footnote{\url{http://jsoc.stanford.edu}} and normalized by exposure time. In addition to selected AIA EUV channels used for context, we focused on the 1600~Å channel for a more detailed analysis. This channel captures C~{\sc iv} and continuum emissions from the upper photosphere and transition region, highlighting flare ribbons and kernels, which often correspond to regions where the flare-accelerated high-energy non-thermal electrons deposit their energy.

To analyze these UV emissions, AIA 1600~Å images were differentially rotated to 19:59:50~UT and base difference images were created by subtracting the reference frame from that time from subsequent images. The $2\times2$ pixels were binned to increase signal-to-noise ratio and to match the pixel scale of the HMI magnetograms, which did not result in significant loss of spatial detail in our analysis. All visualizations, reprojections, and data analysis were performed using the SunPy library.

\subsection{GONG H$\alpha$ filtergrams}

We used full-disk H$\alpha$ filtergrams from the GONG network \citep{Harvey1996} to complement SDO/AIA images in visualizing the filament eruption and flare. These data were used solely for visualization purposes, with no additional analysis performed.

\section{Results}\label{sec:results}

\subsection{STIX HXR pulsations}\label{sec:results_STIX_pulsations}

\begin{figure*}
\centering
  \includegraphics[width=11cm]{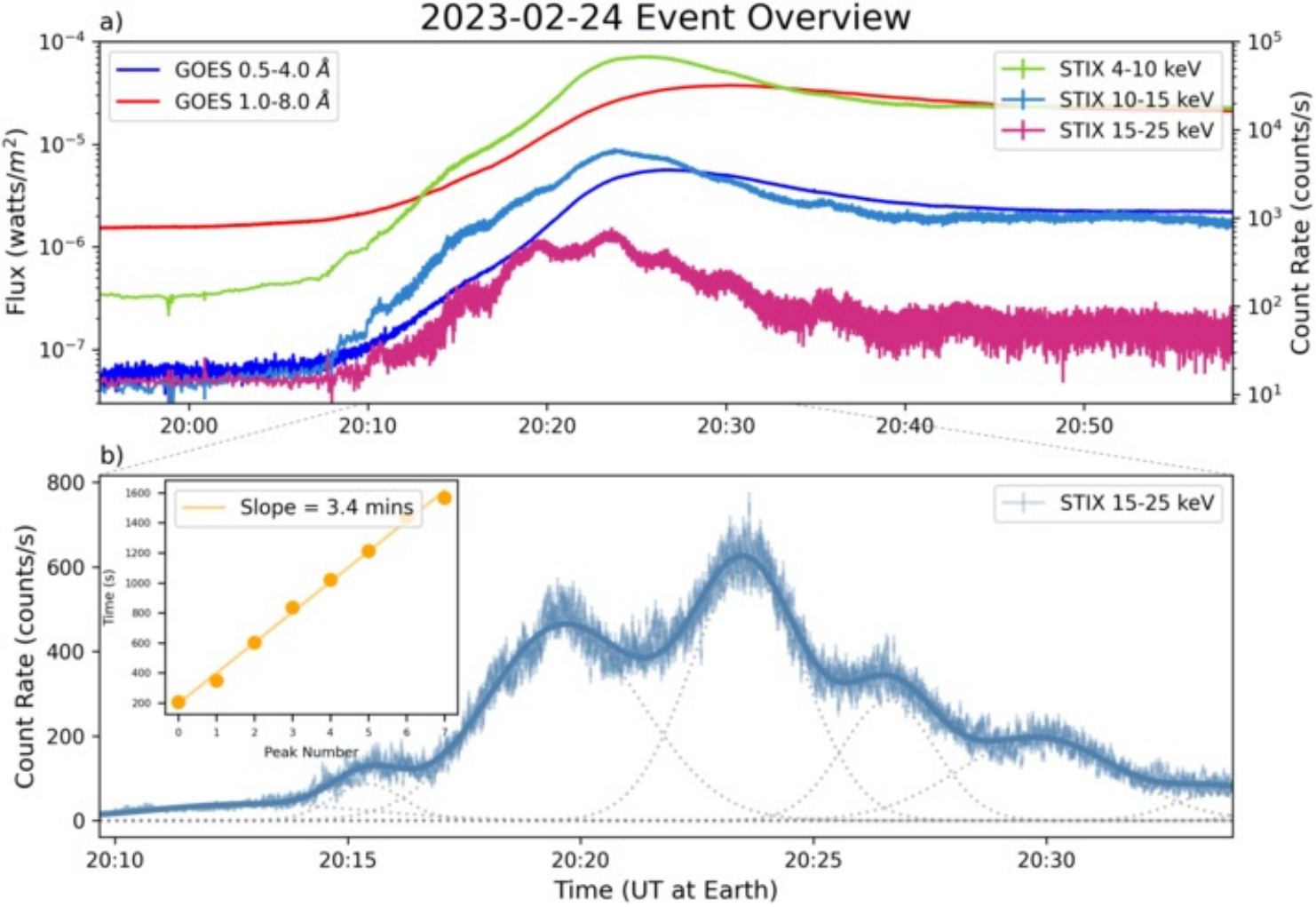}
  \includegraphics[width=7cm]{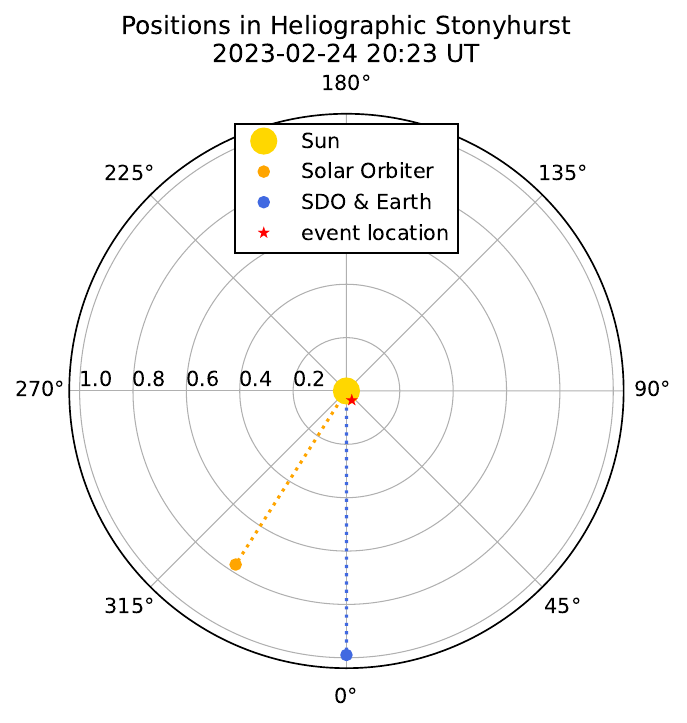}
    \caption{Event overview. Panel a: GOES/XRS and STIX X-ray light curves in multiple energy bands during the flare. Panel b: Shorter time interval, focusing on the pulsations in the STIX 15-25~keV energy band (blue). The observations are fitted by the linear combination (blue) of a series of individual Gaussians (dotted, gray). Right: Spacecraft positions during the event in heliographic stonyhurst coordinates. The position of the main flare loop arcade is marked (red start).}
    \label{fig:event_overview}
\end{figure*}

\begin{figure*}
    \centering
    \includegraphics[width=18cm]{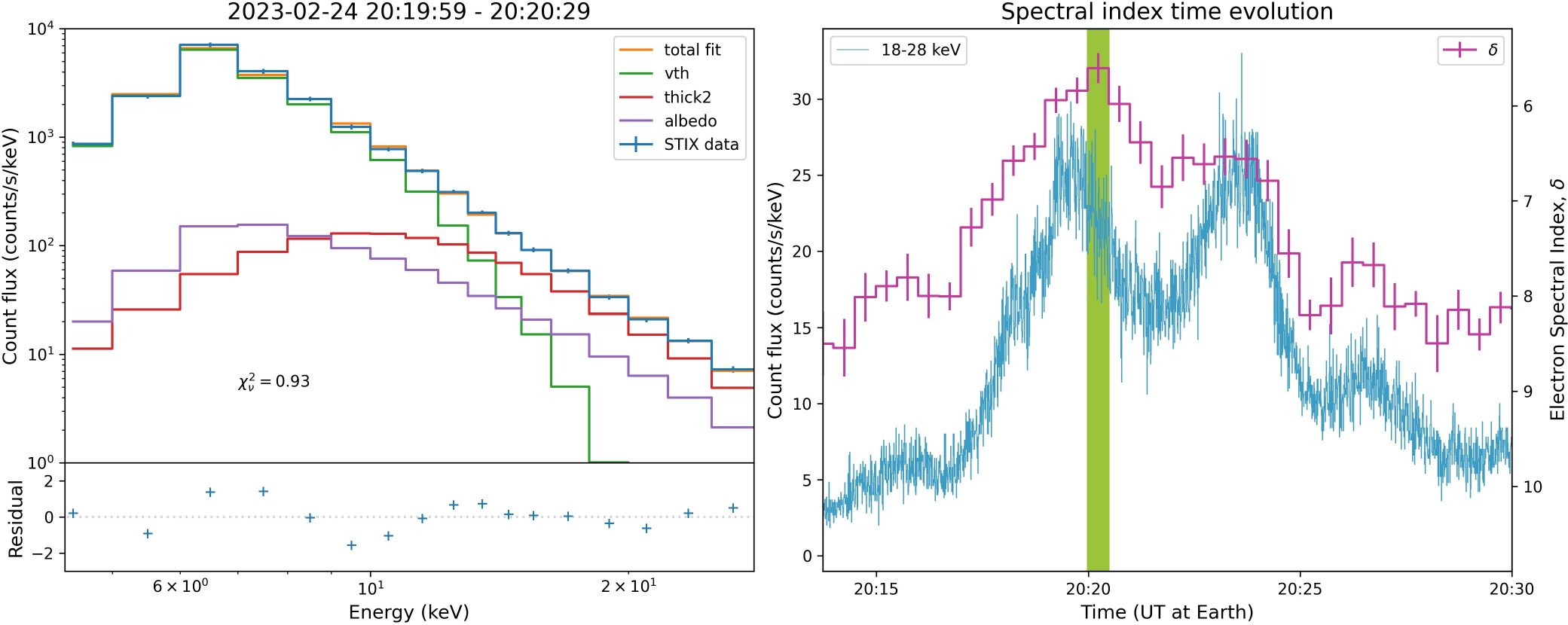}
      \caption{Left: STIX spectrum (blue) fitted with a thermal (vth; green) and a non-thermal (thick2; red) model and an albedo component. Right: Time evolution of the electron spectral index ($\delta$; magenta) derived from a sequence of STIX spectra binned to 30 second intervals and counts in the STIX 18-28~keV energy band (blue). Time interval corresponding to the spectrum (left panel) is marked.}
  \label{fig:STIX_spectral_evolution}
\end{figure*}

An overview of the flare under study in presented in Fig.~\ref{fig:event_overview}.  Solar Orbiter was located at 0.77~AU from the Sun and $33^\circ$ east of the Sun-Earth line, as shown in the right-hand panel. Panel a shows the light curves from the GOES X-ray Sensor (XRS) and STIX in several energy bands over the entire duration of the flare. The flare peaked at a GOES class of M3.7 at 20:30 UT. The STIX light curves display a similar profile to those of GOES/XRS in the lower energy bands; however, the higher energy bands show time-varying signatures (or QPPs) visible throughout much of the flare and mainly identified in the 15-25~keV (magenta) energy band.
In panel b, we focus on this energy band and show only a shorter time interval containing the strongest pulsations. To quantify the pulsation timescale, we fit the pulsations with a linear combination of Gaussians \citep{Collier2023}. A linear fit to the peak times of all the Gaussians gives a pulsation period of approximately 3.4~min.

To further investigate the non-thermal electrons during these pulsations, we fit a sequence of 30~s integrated time bins of the X-ray spectrum observed by STIX to characterize the evolution of non-thermal electrons during the pulsations. Figure \ref{fig:STIX_spectral_evolution} shows an example spectral fit during one interval and the full time evolution of the derived electron spectral index. The high-energy part of the thick target model (thick2) follows a power law whose steepness is given by the electron spectral index ($\delta$). A low spectral index results in a flat spectrum, commonly referred to as a “hard” spectrum, as opposed to a steep, “soft” spectrum with a high spectral index. A lower index means that a greater proportion of the total electrons have been accelerated to high energies, thereby indicating an efficient electron acceleration process.

The comparison of the time evolution of the electron spectral index (Fig. \ref{fig:STIX_spectral_evolution}, right panel; magenta) with the counts detected by STIX in the 18–28 keV energy band (blue) reveals that the spectral index is lowest during HXR peaks and increases between individual pulsations. This behavior, known as soft-hard-soft (SHS) evolution, is a well-documented feature of impulsive X-ray flare emission \citep[e.g.,][]{Grigis2004, Holman2011, Collier2024}. It suggests that the electron acceleration process is modulated at the same frequency as the total emission and has maximum efficiency during the peaks of the HXR emission.

\subsection{SDO/AIA and GONG H$\alpha$ overview}\label{sec:results_SDO_GONG_overview}

\begin{figure*}
\centering
  \includegraphics[width=18cm]{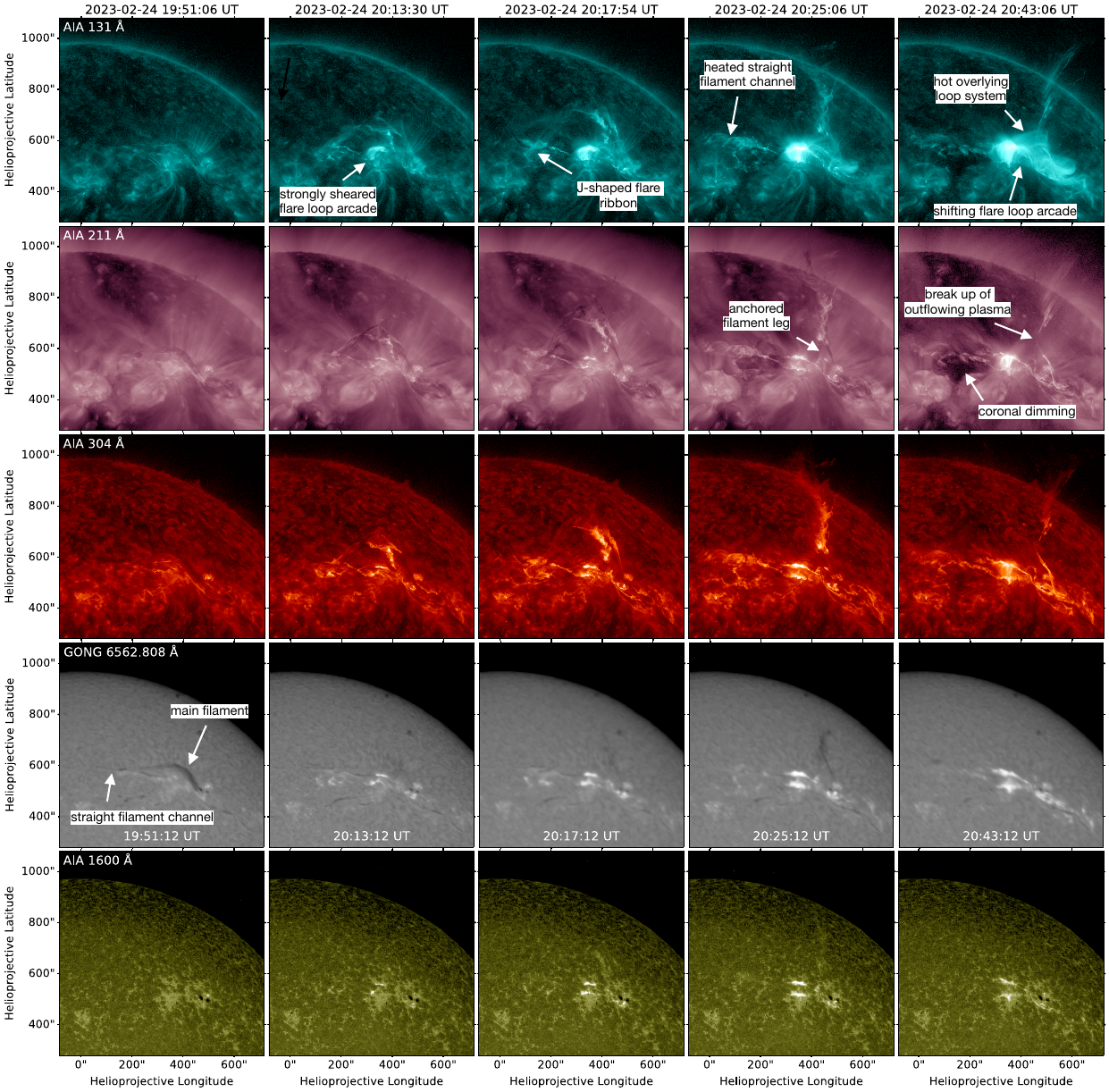}
    \caption{Image sequence of the event in selected SDO/AIA wavelength channels and GONG H$\alpha$ filtergrams. Important features of the event are labeled. The time of the AIA observations is given in the top of each column, while the time of the GONG observations is given separately in each  H$\alpha$ image. The associated movie is available online.
    }
    \label{fig:SDO_GONG_overview}
\end{figure*}

In Fig.~\ref{fig:SDO_GONG_overview}, we present an overview of the spatial evolution of the flare and its associated filament eruption, as observed in selected SDO/AIA wavelength channels and GONG~H$\alpha$ filtergrams. The figure focuses mainly on the time when the HXR pulsations were observed. However, the accompanying movie covers a much longer period. The interesting features discussed in this section are labeled on the images.

The first column shows the filament while it slowly rises and destabilizes. The main part of the filament can be seen as an inverse-S-shaped dark structure in all AIA EUV channels and GONG~H$\alpha$. In addition, there is another straight filament channel extending further east, which is best seen in the GONG H$\alpha$ image.

The images in the second column were taken about 2~min before the peak time of HXR pulsation 2. The AIA 131~Å image shows a strongly sheared flare loop arcade, which then undergoes a strong-to-weak shear transition captured by the following images of the sequence. During this time, the northern flare ribbon expands eastward and the southern ribbon expands westward (see AIA 1600~Å). Later in the flare, the northern footpoints of the loop arcade shift westward, in the opposite direction of the initial ribbon expansion. The resulting flare loops can be seen in the last AIA 131~Å image and at later times in the accompanying movie in SDO/AIA channels sensitive to the cooled flare plasma (e.g., AIA 171 and 193~Å).

The filament eruption is strongly asymmetric. The eastern part escapes rather freely. A J-shaped extension of the main southern flare ribbon forms (e.g., AIA 131~Å image in the third column), which surrounds a coronal dimming region (e.g., last AIA 211~Å image), indicating that the filament is anchored in this region \citep{Sterling1997,Veronig2019}. In contrast, the western part of the filament remains anchored throughout the time we observe the HXR pulsations (e.g., fourth AIA 211~Å image). We observe a continuous outflow of filament plasma during this time. In addition, an overlying large-scale loops system becomes bright in the AIA 131~Å filter (last snapshot in Fig.~\ref{fig:SDO_GONG_overview}), indicating that it heated up. This magnetic loop system seems to be responsible to hold down the anchored western part of the erupting filament.

The last column of images in Fig.~\ref{fig:SDO_GONG_overview} corresponds approximately to the time when the HXR pulsations stop, and captures the moment when the outflow of the filament plasma along the western leg of the filament breaks up (last AIA 211~Å image). The lower part of the plasma falls back towards the Sun, while the upper part escapes.

\subsection{AIA 1600~Å analysis}\label{sec:results_AIA1600}

In this section, we present the results of our analysis of the temporal and spatial evolution of UV brightenings observed in an AIA 1600~Å base-difference image sequence (see section \ref{sec:methods_AIA_HMI}) and their spatiotemporal relation to the STIX HXR pulsations.

\subsubsection{UV light curves from different subregions}\label{sec:results_AIA1600_light_curves}

\begin{figure*}
\centering
  \includegraphics[width=18cm]{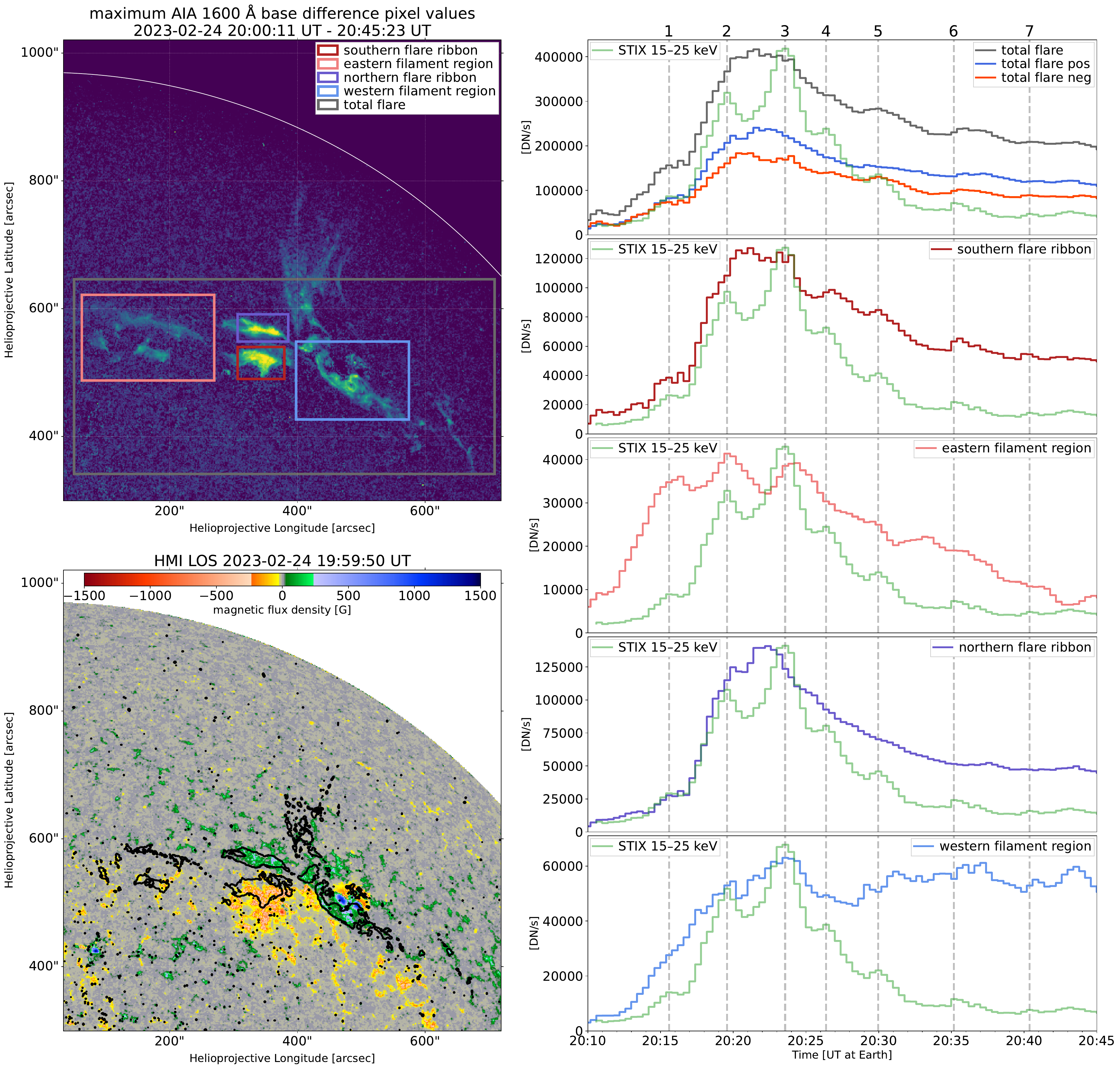}
    \caption{Overview of AIA 1600~Å base difference light curves in different subregions. Top left: Map showing the maximum pixel values in AIA 1600~Å base difference maps during the given time interval with four subregions marked. Bottom left: HMI LOS magnetogram with contours (25 DN/s) of the maximum AIA~1600~Å brightness map above. Right: Light curves showing the time evolution of the total AIA 1600~Å base difference counts in each of the marked subregions. The emission in the total flare mask shown in the top panel is further divided into emission from pixels corresponding to a positive or negative polarity region, as derived from the shown HMI LOS magnetogram. The STIX 15–25 keV
    light curve (green) is shown as a reference, and peak times of the HXR pulsations are marked by dashed lines and numbered at the top.}
    \label{fig:AIA1600_submaps_and_lightcurves}
\end{figure*}

As a first step, we analyzed the time evolution of the spatially integrated AIA 1600~Å emission from different subregions of the flare. Figure \ref{fig:AIA1600_submaps_and_lightcurves} shows the extent of all subregions (total flare region, southern flare ribbon, eastern filament region, northern flare ribbon, western filament region), the relation of all flaring regions to the HMI LOS magnetic field, and the extracted light curve from each region in comparison with the STIX HXR pulsations.

We find that the light curve of the total flare indeed shows a good correlation with most HXR pulsations, except for a strongly enhanced emission between HXR pulsations 2 and 3. Interestingly, when this emission is split according to the magnetic polarity in the HMI LOS magnetogram, the UV response to the HXR pulsations is generally much stronger in negative polarity regions. 
The emission from positive polarity regions shows some correlation, but it is much less prominent.

The contours of the AIA 1600~Å emission plotted on the HMI LOS magnetogram show which parts of the flare belong to each polarity. The southern flare ribbon and its J-shaped extension into the eastern filament region are located in the negative polarity, while the northern flare ribbon and its extension into the western filament region are located in the positive polarity. The pre-eruption filament followed the inverse-S-shaped polarity inversion line between the two flare ribbons. This illustrates an asymmetry between the flare ribbons in both polarities. While the compact southern flare ribbon is located on the inner side of the bend of the PIL, the narrower northern flare ribbon follows the outer side of the bend as it continues into the western filament region. This is consistent with the observation of the flare loop arcade fanning out to the northwest in the SDO/AIA observations (see Fig.~\ref{fig:SDO_GONG_overview}).

We also show the light curves in each of the four marked subareas in Fig.~\ref{fig:AIA1600_submaps_and_lightcurves}.
The southern flare ribbon shows a very good correlation with all HXR pulsations, with only some excess emission after the second HXR peak, resulting in a delayed peak time of the UV emission.
The eastern filament region shows a surprisingly strong correlation with HXR pulsations 1 to 3, and perhaps even some response to HXR pulsations 4 and 5.

The AIA 1600~Å UV light curve from the northern flare ribbon seems to closely follow the STIX HXR pulsations 1 and 2. The UV emission peaks between HXR peaks 2 and 3 and then decreases continuously with no visible response to any of the other HXR pulsations.
AIA 1600~Å emission from the western filament region shows peaks in correlation with HXR pulsations 2, 3, and perhaps 4. Then there is a general increase in UV emission in this region just before HXR peak 5, resulting in the noisy plateau in the light curve in the second half of the time interval analyzed.

\subsubsection{General evolution of UV sources and derived parameters}\label{sec:results_AIA1600_masks_and_parameters}

We analyzed the overall spatial evolution of the AIA 1600~Å emission (e.g., the expansion of the flare ribbons) and derived parameters from this evolution. The analysis again focuses on each of the four subregions introduced in Fig.~\ref{fig:AIA1600_submaps_and_lightcurves}.
To track the emission enhancements over time, we set an individual fixed threshold for each subregion, depending on the magnitude of the brightness increase in each region. We then create several masks based on the pixel counts in an AIA 1600~Å base-difference image sequence.
The different masks for a given time are defined as follows:
the instantaneous mask contains any pixel that exceeds the set threshold at that time.
The new mask contains any pixel that exceeds the threshold for the first time at that time.
The cumulative mask contains any pixel that has exceeded the threshold up to and including that time.
From the mask sequences, we derive the time evolution of the following parameters:

\begin{itemize}
    \item $\mathrm{A_{inst}}$: the area within the instantaneous mask;
    \item $\mathrm{A_{cumu}}$: the area within the cumulative mask;
    \item $\Phi_{+}$ or $\Phi_{-}$: The total negative or positive flux in HMI LOS magnetogram pixels within the cumulative mask, depending on the main polarity of the subarea;
    \item B: The mean magnetic flux density in the new mask;
    \item COM$_{\mathrm{x}}$ and COM$_{\mathrm{y}}$: The x and y coordinates of the center of mass within the instantaneous mask;
\end{itemize}

\begin{figure}
  \resizebox{\hsize}{!}{\includegraphics{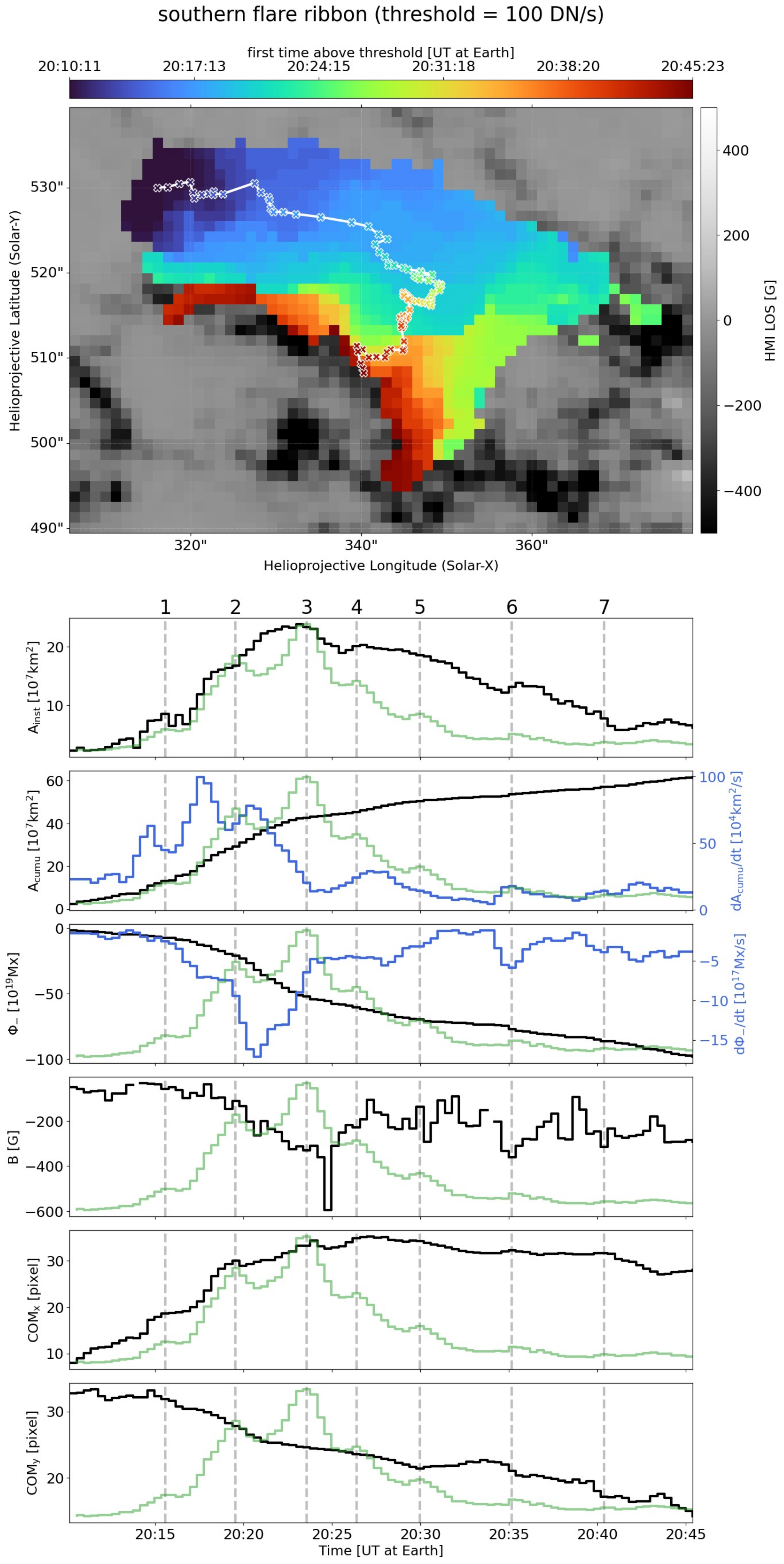}}
  \caption{Spatiotemporal evolution of the AIA 1600~Å emission in the southern flare ribbon subarea introduced in Fig.~\ref{fig:AIA1600_submaps_and_lightcurves} and derived parameters. Top panel: Pixels are colored according to the time at which they first exceeded the set threshold (100 DN/S) during the time interval indicated on the color bar. The total extent of the colored area corresponds to the final cumulative mask. The background shows the magnetic flux density according to the HMI LOS magnetogram recorded at the beginning of the time interval. The center of mass (COM) within the instantaneous mask of each time step is marked (X). Bottom panels: Time series of parameters derived from the mask evolution and the HMI LOS magnetogram shown. From top to bottom: The total area of the instantaneous mask ($\mathrm{A_{inst}}$), the total area of the cumulative mask ($\mathrm{A_{cumu}}$), the total flux within the cumulative mask ($\Phi_{+}$ or $\Phi_{-}$ depending on the main polarity of the subregion), the mean magnetic flux density in newly added pixels (B), and the x and y coordinates of the COM. Smoothed derivatives of selected parameters are shown in blue. The STIX 15–25 keV light curve (green) is shown as a reference, and peak times of the HXR pulsations are marked by dashed lines and numbered at the top. Associated movie is available online.
  }
  \label{fig:parameters_southern_ribbon}
\end{figure}

\begin{figure}
  \resizebox{\hsize}{!}{\includegraphics{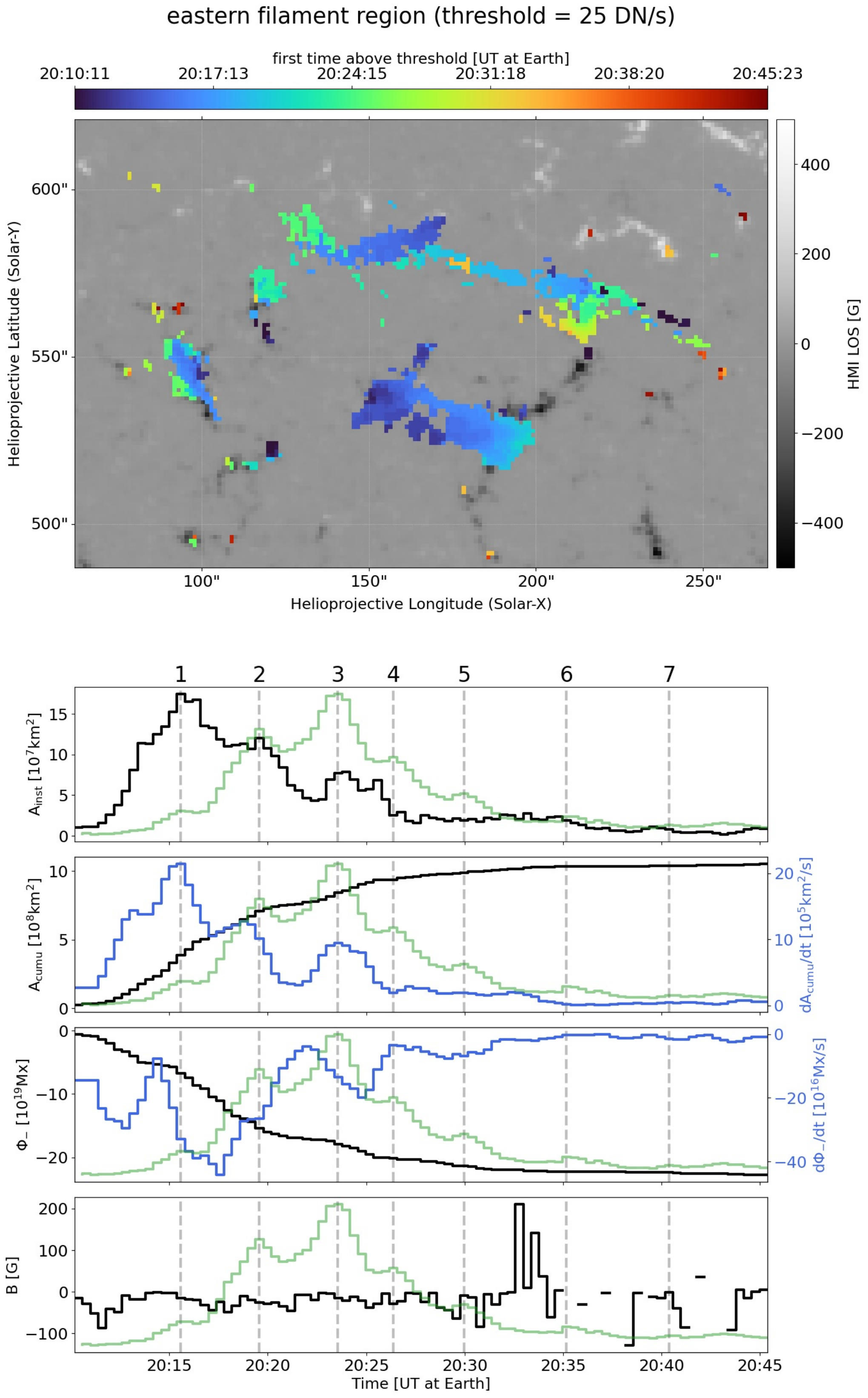}}
  \caption{Same as Fig.~\ref{fig:parameters_southern_ribbon} but for the eastern filament region. The COM location and the time evolution of its x- and y-coordinate are not shown for this region. Associated movie is available online.
  }
  \label{fig:parameters_eastern_filament}
\end{figure}

\begin{figure}
  \resizebox{\hsize}{!}{\includegraphics{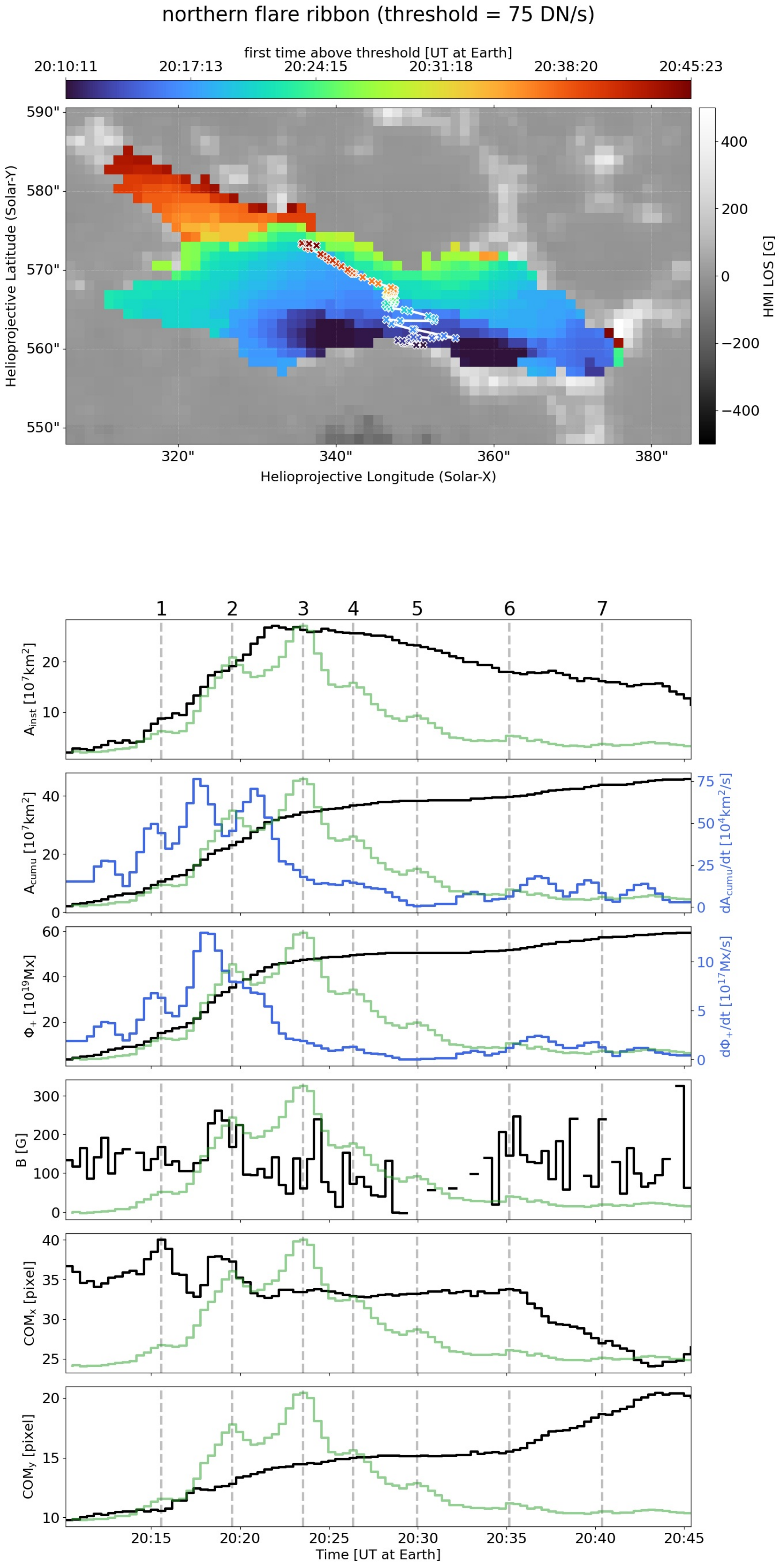}}
  \caption{Same as Fig.~\ref{fig:parameters_southern_ribbon}, but for the northern flare ribbon. Associated movie is available online.
  }
  \label{fig:parameters_northern_ribbon}
\end{figure}

\begin{figure}
  \resizebox{\hsize}{!}{\includegraphics{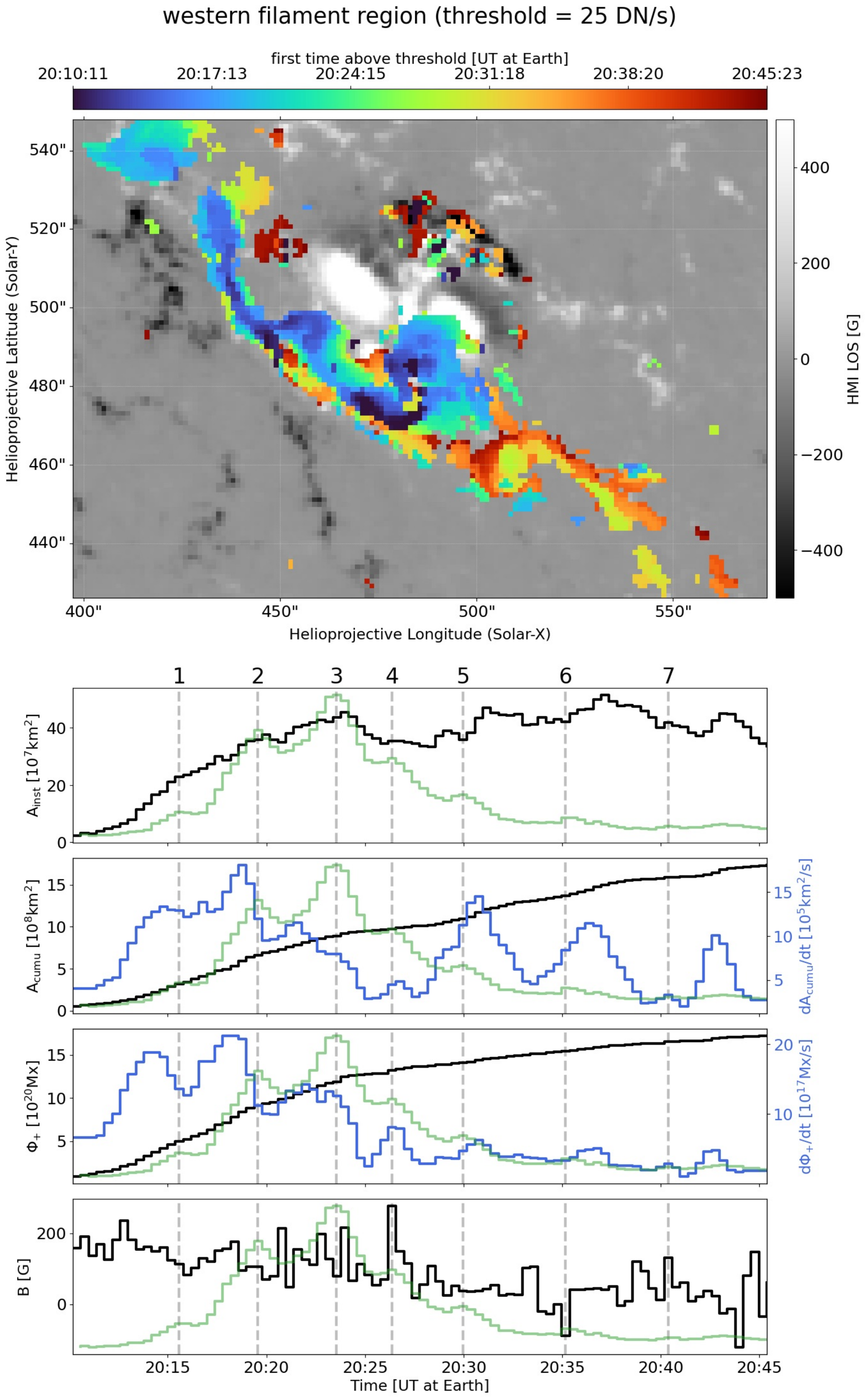}}
  \caption{Same as Fig.~\ref{fig:parameters_southern_ribbon}, but for the western filament region. The COM location and the time evolution of its x- and y-coordinate are not shown for this region. Associated movie is available online.
  }
  \label{fig:parameters_western_filament}
\end{figure}

Figures \ref{fig:parameters_southern_ribbon}, \ref{fig:parameters_eastern_filament}, \ref{fig:parameters_northern_ribbon}, and \ref{fig:parameters_western_filament} show the final extent of the cumulative mask for the four subregions after the investigated time period (20:10:11 to 20:45:23~UT) with pixels colored according to the time they first exceeded the threshold (new mask).
In addition, the center of mass (COM) of pixels in each instantaneous mask is calculated for the southern and northern flare ribbons and shown in the respective plots to give a general impression of their expansion. 
Each figure is accompanied by a movie showing the evolution of the instantaneous and cumulative masks and the COM (only for the main flare ribbons) in relation to the original AIA 1600~Å images, the base difference version, and the HMI LOS magnetogram.
We will analyze this motion in greater detail later in Section \ref{sec:results_AIA1600_time_dist}.

For the southern flare ribbon, the threshold was set to 100~DN/s in the base difference images.
The mask sequence shown in Fig.~\ref{fig:parameters_southern_ribbon} and the accompanying movie highlight the general motion of the southern ribbon. First, it expands rapidly from east to west during HXR pulsations 1 and 2 (blue area). This is followed by a southeastward expansion (green to red) that occurs separately in two regions, namely at the westernmost part of the ribbon and near its center. The COM first follows the mostly westward expansion and then moves southeastward between these two regions. This evolution is also visible in the time series of the COM$_{\mathrm{x}}$ coordinate. In addition to this overall evolution, the COM temporarily moves westward with each pulsation. This indicates that the AIA 1600~Å emission dominantly increases in the western part of the ribbon relative to the COM during each pulsation.

The time evolution of the instantaneous area follows a smooth profile that shows some response to the HXR pulsations, but is generally weak and inconsistent. The derivatives of their time evolution are not well correlated with the pulsations, except for HXR peak 6.
These observations would suggest a smooth expansion of the ribbon, with only small variations caused by the pulsations. However, the exact evolution of these parameters is of course sensitive to the threshold used, and in particular the smooth evolution of the ribbon area is mainly a result of a rather low threshold. In Fig.~\ref{fig:parameters_southern_ribbon_higher_threshold} in the appendix, we show the area and total flux evolution extracted with a higher threshold (300 DN/s). This shows a stronger correlation of the instantaneous area with the pulsations, and also a better correlation of the derivatives of the cumulative area and magnetic flux, especially during later HXR peaks. This suggests that the ribbon evolution may be more modulated by the pulsation than the low threshold suggests. However, the low threshold better captures the full evolution of the ribbon and is therefore shown here.

The magnetic flux density in the newly added pixels decreases from its initial level of about -50~G during the first HXR peak to a minimum of about -300~G during the third HXR peak. The spike just after this is caused by the addition of a very small area that happens to be in a high magnetic flux region. The flux density starts to decrease again after the third HXR peak and then becomes quite noisy due to the small area growth rate. There is no indication of change in the magnetic flux density on the same time scales as the pulsations.

For the eastern filament region (Fig.~\ref{fig:parameters_eastern_filament}), we set the threshold to 25~DN/s, due to the much fainter emission increase in this region compared to the main part of the ribbon.
The mask evolution shows two regions with very different behaviors.
First, there is a kernel in the south that starts at about x = 150", y = 540" and then moves mostly westward.
The other area is the emission along the J-shaped ribbon, which is mostly added to the cumulative mask in patches and does not show a continuous evolution. The accompanying movie shows that the emission from this region seems to be dominated by plasma flowing into the second straight filament channel from the direction of the main flare region (plasma motion from west to east). The derived parameters indicate that both the instantaneous area and the derivative of the cumulative area seem to be strongly correlated with the first three HXR pulsations and have a very similar profile. This means that most of the pixels brightening above the threshold are completely new pixels and are fully added to the cumulative area.

The northern flare ribbon (Fig.~\ref{fig:parameters_northern_ribbon}) was analyzed with a threshold of 75~DN/s.
This flare ribbon starts as two separate kernels and then expands mostly to the northeast.
The instantaneous area may show some correlation in the beginning, but loses all correlation after the second HXR peak, similar to the light curve in Fig.~\ref{fig:AIA1600_submaps_and_lightcurves}.
The magnetic flux density peaks just before the second HXR peak to about 250~G, but then quickly becomes unreliable due to the low area growth rate of this ribbon during the later HXR peaks. Similar to the southern flare ribbon, no correlation of the magnetic flux density with the pulsations is found.
Also, the COM$_{\mathrm{x}}$ component pulsates westward during the first two HXR peaks, indicating a stronger AIA 1600~Å response to the HXR pulsations west of the COM location.
The COM then remains almost stationary until HXR peak 6, after which it continues on a straight path to the northeast as a second expansion of the ribbon begins, visible as the orange to red pixels in the top panel.

Finally, we show the results of the western filament region derived with a threshold of 25~DN/s in Fig.~\ref{fig:parameters_western_filament}.
At first glance, the evolution of this region looks much more complex.
However, it basically consists of a thin ribbon (blue regions) running from the upper left corner of the image along the positive side of the PIL, with a hook extending towards the sunspots. This hook then twists during the flare and the thin ribbon expands slightly.
Just before HXR peak 5 there is an impulsive general enhancement of this area, activating the additional areas colored in orange to red, which remain active throughout HXR peaks 5, 6, and 7, and is also visible as the general enhancement of the instantaneous area during this later phase. Interestingly, the derivative of the cumulative area seems to show some pulsations, with three strong peaks during the second half of the time series that seem to have a longer period than the HXR pulsations. The derivative of the total positive magnetic flux also seems to have a general time-varying and pulsating behavior throughout the flare.

\subsubsection{Detailed spatial and temporal evolution of UV sources}\label{sec:results_AIA1600_time_dist}

\begin{figure}
  \resizebox{\hsize}{!}{\includegraphics{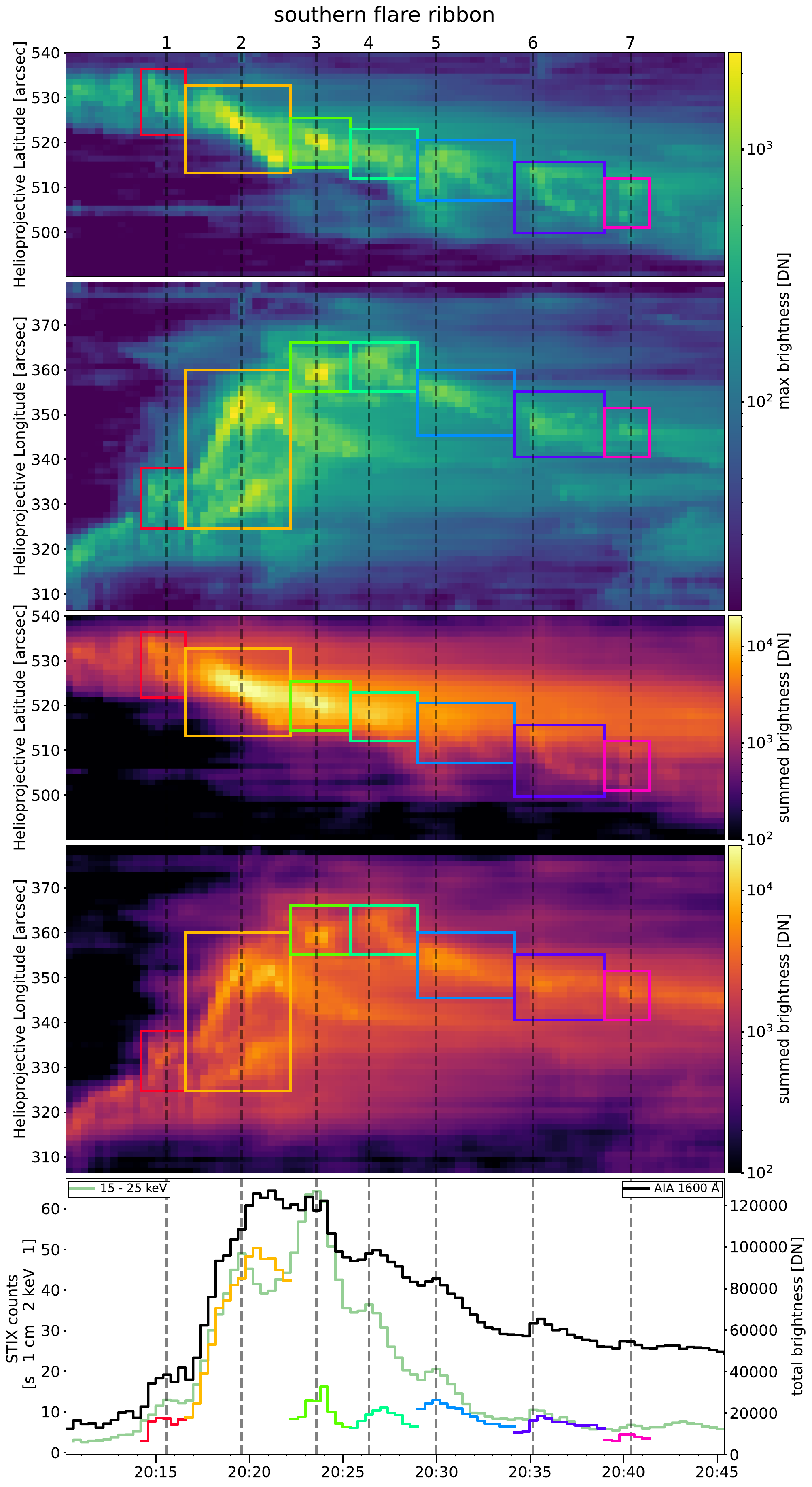}}
  \caption{Time-distance plots of the maximum/summed pixel values along rows/columns of a AIA 1600~Å base difference image sequence of the southern flare ribbon. Upper panel: The maximum pixel values along each row as a function of time. Colored boxes indicate the extent and duration of masks placed around each peak. Second panel: Same, but for the maxima along each column. Third panel: The sum of pixel values along each row. Fourth panel: Sum along the columns of the image. Bottom panel: Time series of the total AIA 1600~Å counts in the entire subarea (black). Colored time series correspond to the counts within the matching masks drawn on the time-distance plots above. The STIX 10-15~keV light curve (green) is plotted for comparison, and HXR peaks are marked with dashed lines throughout this plot and marked at the top of the figure. The associated movie visualizes how the time-distance plots of the maximum values are obtained and how their evolution relates to the images. Associated movie is available online.
  }
  \label{fig:time_dist_southern_ribbon}
\end{figure}

\begin{figure}
  \resizebox{\hsize}{!}{\includegraphics{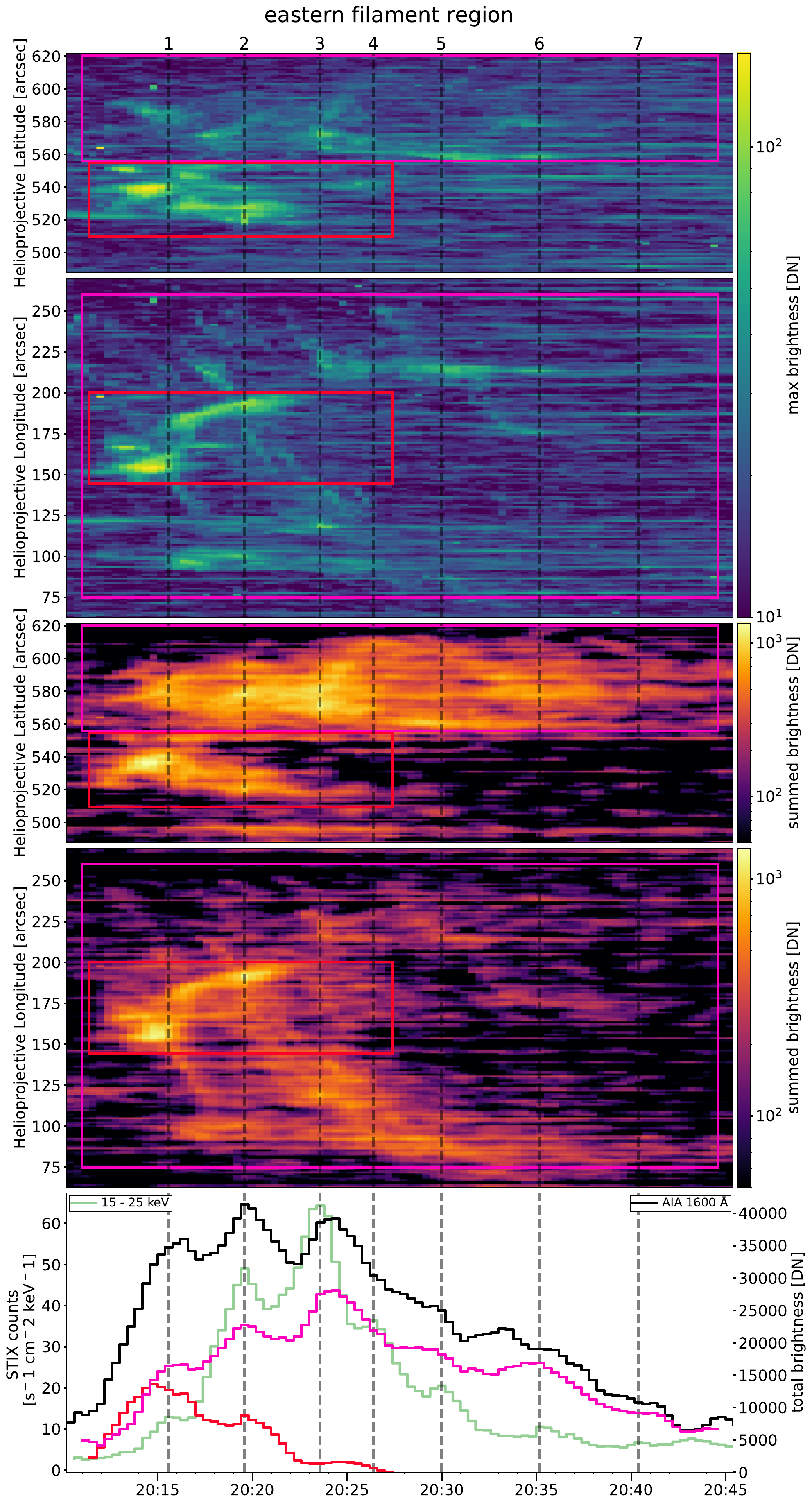}}
  \caption{Same as Fig.~\ref{fig:time_dist_southern_ribbon}, but for the eastern filament region. Associated movie is available online.
  }
  \label{fig:time_dist_eastern_filament}
\end{figure}

\begin{figure}
  \resizebox{\hsize}{!}{\includegraphics{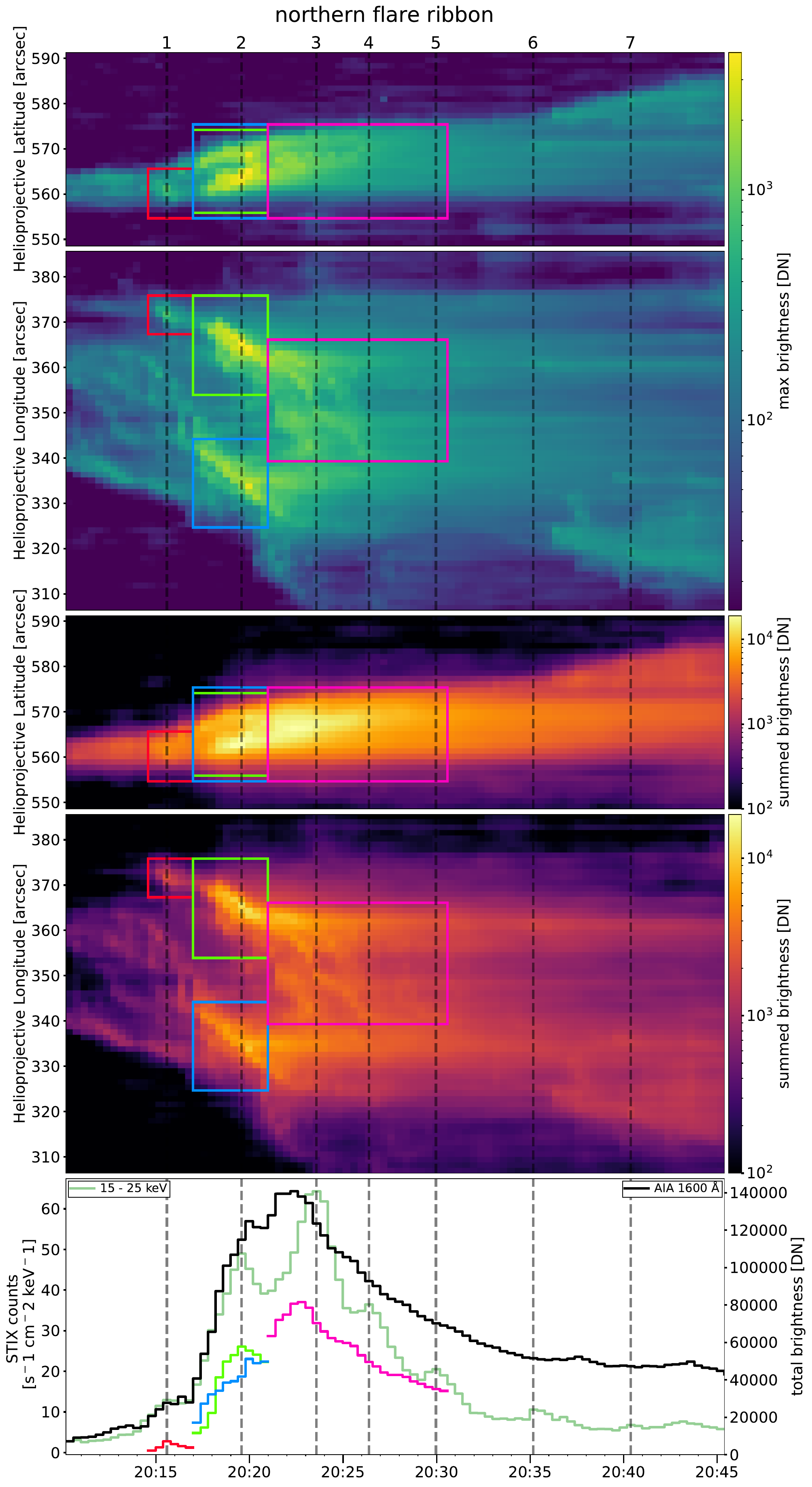}}
  \caption{Same as Fig.~\ref{fig:time_dist_southern_ribbon}, but for the northern flare ribbon. Associated movie is available online.
  }
  \label{fig:time_dist_northern_ribbon}
\end{figure}

\begin{figure}
  \resizebox{\hsize}{!}{\includegraphics{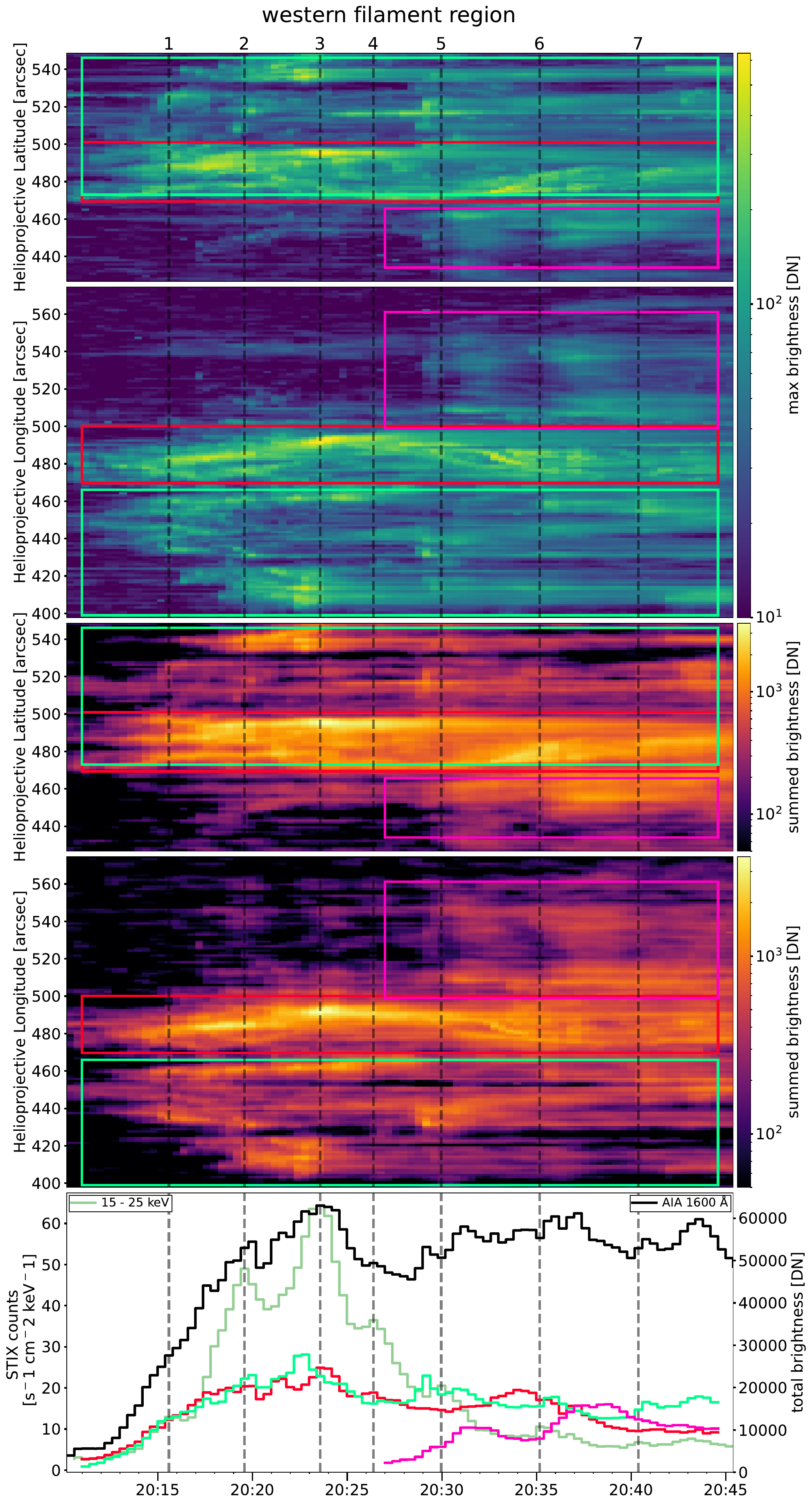}}
  \caption{Same as Fig.~\ref{fig:time_dist_southern_ribbon}, but for the western filament region. Associated movie is available online.
  }
  \label{fig:time_dist_western_filament}
\end{figure}

In addition to analyzing the overall evolution of each region, we performed an in-depth analysis of the spatiotemporal evolution of the AIA 1600~Å emission in correlation with the HXR pulsations, with the goal of pinpointing the source of the emission and quantifying its brightness evolution and motion. We achieved this by producing time-distance plots for each subregion, which are shown in Figures \ref{fig:time_dist_southern_ribbon}, \ref{fig:time_dist_eastern_filament}, \ref{fig:time_dist_northern_ribbon}, and \ref{fig:time_dist_western_filament}.

They look similar to the classic j-maps used to track, for instance, filament eruptions, which require the placement of a fixed slit in the image sequence. The problem is that this method only captures the motion along this slit, which is not sufficient to capture the often quite complex motion of flare kernels, which can move in two dimensions.
Our method instead encodes the full motion of all flare kernels into two time-distance plots: one for the motion in longitude and one for the motion in latitude.
For the first row in the plots, we find the maximum pixel value along each row of a base difference image. We obtain a vertical array containing the maximum pixel values along each row of the current base-difference image as a function of the row's latitude. We extracted such a vertical array from each base-difference image taken in the analyzed time interval and then stacked them as a function of time. The resulting image encodes the position and motion of the brightest sources (flare kernels) along the latitude as a function of time.
The second row of the plot is made using the same method, except that the maximum value along each column of an image is used. This results in a horizontal array showing the distribution of maxima along the longitude. The horizontal arrays from the entire image sequence are again stacked in time, and the resulting image is rotated 90$^\circ$ to fit better into the figures. Together, the two time-distance plots encode the full motion of the flare kernels in two dimensions and through time. The movie accompanying each figure shows how each horizontal and vertical array of maxima is obtained and how they are stacked in time.

In many cases, the position of relevant flare kernels is immediately obvious from these maximum intensity stacks, since this method is very sensitive to bright compact sources. Once we have localized the position of relevant kernels, we can use both time-distance plots to define the boundaries (longitude, latitude, and time) of a mask that we want to place around a particular area to extract its light curve.
The boundaries of these masks are shown in the time-distance panels, and the extracted light curves are shown in the bottom panel and compared to the time series of the total AIA 1600~Å light curve from the whole region and the STIX HXR light curve. The same masks are also shown in the movie associated with each figure, which helps to locate the pulsating kernels in the AIA 1600~Å base-difference movie.

Rows three and four of each plot use the same principles, but instead of showing the maximum pixel values, they show the sum of the pixel values along each row or column of the image. In principle, this is the same as a classic J-map, which places a horizontal or vertical slit across the entire image.
The advantage of the sum is that it preserves information about the total brightness along a row/column, while the maximum captures only the brightest source, but does not include any additional fainter emission. It is therefore suitable for capturing emission spread over a larger area and gives a better impression of the relative brightness of different areas.
However, the sum tends to obscure the true position and motion of flare kernels, since even though they may be the brightest pixels, they often represent only a fraction of the total emission along the entire flare ribbon. The advantages and disadvantages of both methods will become clearer as we discuss the four areas of interest below.

For the southern flare ribbon (Fig.~\ref{fig:time_dist_southern_ribbon}), the time-distance plots reveal the complex evolution of multiple flare kernels, with the strongest motions appearing along the longitudinal direction (panels two and four).
The ribbon starts as a single kernel in the east ($x = 320$~arcsec) and then begins to expand slight westward.
The response to the first pulsation occurs after this initial expansion at about $x = 330$~arcsec (red box). This is followed by the much more rapid westward motion (with a speed of about 80~km/s) of a compact, continuously brightening kernel that leads to the second HXR peak (upper part of the orange box).
During the second HXR peak, another kernel seems to appear right next to it ($x = 350$~arcsec), which first follows the westward motion for a short time and then turns sharply to move back to the east. 
Another kernel brightens at the same longitude as the first pulsation ($x = 330$~arcsec; lower half of the orange box) and peaks after the second HXR peak. Together, the emission from these multiple kernels active around the time of the second HXR pulsation (orange box) produces a light curve whose peak is shifted back in time relative to the HXR light curve of the second pulsation. However, just the westernmost, fast moving kernel correlates well with the exact peak time of the HXR pulsation.
HXR peak 3 then corresponds to a single, very compact AIA 1600~Å kernel that does not appear to move significantly in longitude (green box).
The westernmost position of flare kernels is reached during HXR peak 4 (cyan box), after which we see a continuous drift of flare kernels back to the east during HXR pulsations 5, 6, and 7 (about 20~km/s).

Looking at the latitude evolution (first panel), the initial kernel of the ribbon remains at a constant latitude ($y = 530$~arcsec) until the first HXR peak (red), after which a southward expansion begins. This expansion accelerates towards the second HXR peak (orange) and then produces an almost linear southward streak (with about 40~km/s). The summed plot (panel four) shows that this rapid southward motion of the kernel is largely obscured by the main ribbon emission, which lags behind. 
The southward motion of the kernel eventually stops ($y = 515$~arcsec), but we can see a much fainter enhancement continuing southward (reaching about $y = 500$~arcsec).
Comparison with the movie shows that the southwestern outline of the final ribbon area is already activated at this time.
After HXR peak 3 (green) there is another faint southward streak of similar intensity and velocity that fills in part of the ribbon area just east of the first enhancement. Assuming that this streak started at the kernels associated with the third HXR peak, the enhancement moved southward with a speed of about 90 km/s.
The kernel associated with HXR peak 4 (cyan) actually moves into the southern region of the ribbon, resulting in a much brighter streak in the time-distance plot.
The AIA 1600~Å kernels corresponding to later HXR peaks all start in the north and then move south with varying intensities and velocities.
Each pulsation continuously fills another slice of the ribbon area already outlined after the second HXR peak.

To summarize the evolution in this southern ribbon, it appears that the main positions of the kernels correlated with the HXR pulsations are on a continuous path, first moving rapidly westward until the fourth HXR peak, and then slowly drifting back eastward. In addition, each pulsation moves southward or activates southern areas of the ribbon. The brightening associated with the second HXR peak already outlines the edge of the final ribbon area, while subsequent pulsations fill the outlined area from east to west and from north to south.
Additional kernels brightening during the second HXR pulsation fill in the remaining, more northern ribbon area, but do not appear to respond to the following HXR pulsations. These results confirm the two separate paths of the eastward expansion of the ribbon seen in Fig.~\ref{fig:parameters_southern_ribbon} and explain why the COM pulsates westward with each HXR peak.

The time-distance plots of the eastern filament region (Fig.~\ref{fig:time_dist_eastern_filament}) further illustrate the different behavior of the two subregions, already highlighted in the discussion of Fig.~\ref{fig:parameters_eastern_filament}.
In the southern subregion (red box) we observe two compact kernels that dominate the time-distance plot generated from the maximum pixel values (top two panels). The brighter one is very short-lived and is brightest just before the first HXR peak ($x = 150$, $y = 540$), the other one becomes active just after HXR peak 1, stays active and peaks again during HXR peak 2. 
Together, they produce a light curve (red; bottom panel) that roughly correlates with the first two HXR pulsations. 
However, much of the total AIA 1600~Å emission (black) actually comes from the region to the north (pink box), which includes part of the J-shaped ribbon and the additional straight filament channel. The maximum plots in the top two panels of the figure do not capture this emission very well because it is actually spread over a larger region.
However, the total emission and pulsations are visible in the time-distance plots using the summed pixel values over each axis (panels 3 and 4). The light curve (pink; bottom panel) from this subregion shows the very good correlation with the pulsations. Only pulsation 4 seems to be missing a clear peak in the light curve, but it could just be hidden in the decay of the previous peak.
The movie accompanying this figure again shows that much of the emission from this subregion seems to come from plasma injected into the straight filament channel from the direction of the main flare loops.

The results from the northern flare ribbon are shown in Fig.~\ref{fig:time_dist_northern_ribbon}.
It is immediately apparent that the kernel evolution in the northern ribbon is much simpler than in the southern ribbon.
The response to the first HXR peak mainly comes from a very compact kernel (red box).
During the second HXR pulsation, two separate kernels (green and blue) form and move eastward at a similar speed (about 40~km/s), resulting in nearly parallel streaks in the time-distance plots. Both peak in brightness during the second HXR peak and then fade continuously.
During the third HXR pulsation, there is more diffuse emission from the region between the two kernels (pink), which is not strongly correlated with HXR peaks. The integrated emission from this region (pink; bottom panel) peaks at least one minute before HXR peak 3 and also causes the uncorrelated peak in the light curve of the whole region (black). We find no further correlated response of the AIA 1600~Å emission to the remaining HXR peaks in this region.
After HXR peak 6 we see the signature of further expansion of the ribbon to the northeast. This appears to be a completely separate process from the initial expansion and was previously discussed as the red area in Fig.~\ref{fig:parameters_northern_ribbon}.

Finally, the western filament region appears rather complex in the time distance plot (Fig.~\ref{fig:time_dist_western_filament}).
However, the plot can basically be divided into three subregions. The narrow part of the ribbon extending from the northern flare ribbon along the PIL towards the southwest (green), the hook near the sunspots (red), and the region further southwest (violet), which becomes active only around HXR peak 5. Comparison with the accompanying movie makes this separation clearer.
The strongest AIA 1600~Å response occurs during the third HXR peak from a kernel located between the two sunspots (red box), which also shows a strong correlation in the light curve (red; lower panel). In addition, there seems to be a much weaker response from the same kernel to HXR peak 4, but the rest of the light curve from this central region does not seem to correlate with the following HXR pulsations. The green light curve from the narrow ribbon does appear to contain pulsations, but they don't always line up perfectly with the HXR peaks. This region is basically the direct extension of the northern flare ribbon, and also the region where the flare loops begin to connect during the flare (see Fig.~\ref{fig:SDO_GONG_overview} and the accompanying movie). It would therefore make sense for the UV pulsations to follow the expansion of the flare loops from the northern ribbon into this western filament region. The third region (pink) shows a general enhancement around HXR peak 5 in the time-distance plots that persists until the end of the pulsations, and was also discussed as the orange to red areas in Fig.~\ref{fig:parameters_western_filament}. The emission from this region appears to pulsate out of phase with the HXR pulsations, which were also detected as pulsations in the area growth rate in the same figure.

\subsection{STIX HXR images}\label{sec:results_STIX_imaging}

\begin{figure}
  \resizebox{\hsize}{!}{\includegraphics{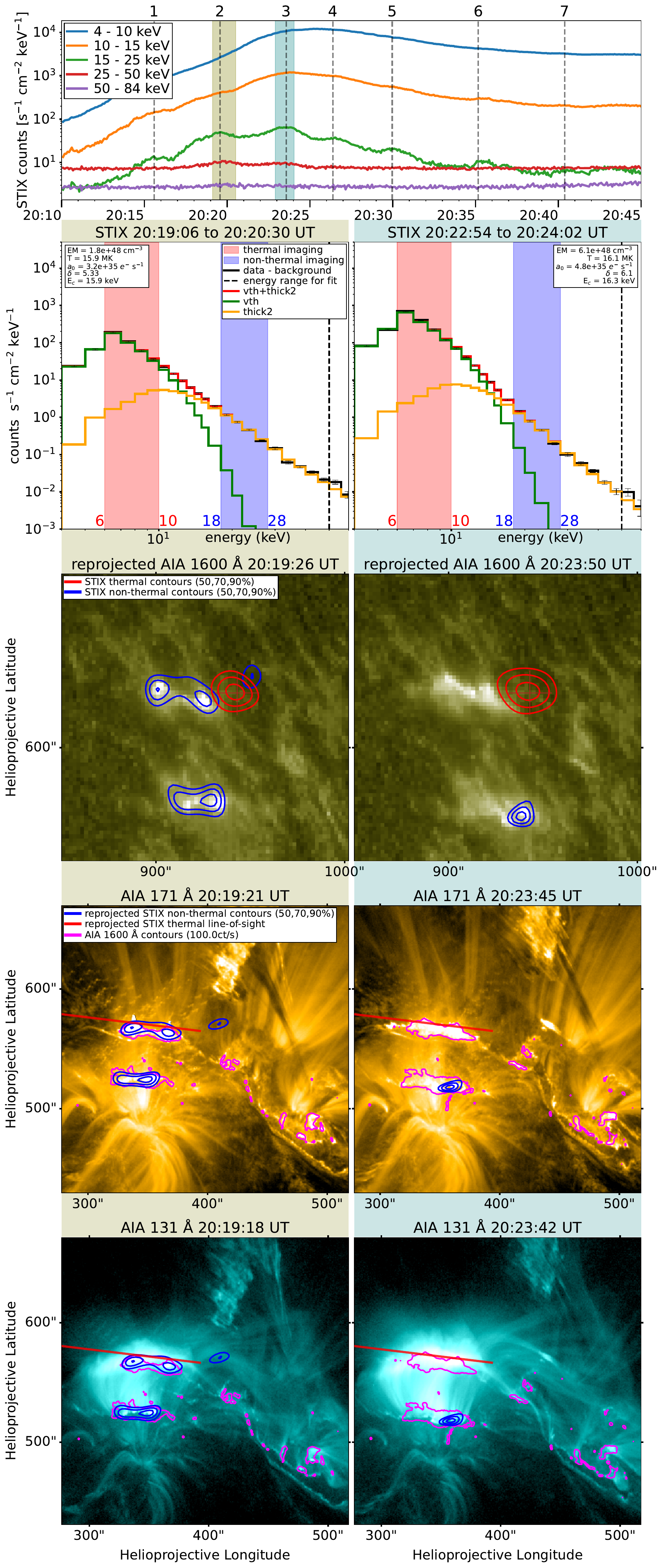}}
  \caption{STIX observations for HXR pulsations 2 and 3. From top to bottom: 1) STIX light curves in different energy bands, with two time intervals marked. 2) STIX spectrum during each time interval fitted with a thermal (vth; green) and non-thermal (thick2; yellow) model. 3) Thermal (red) and non-thermal (blue) mem\_ge image contours corresponding to the energy intervals marked in the spectra, plotted over reprojected AIA 1600~Å images. 4 and 5) The same STIX sources reprojected to the AIA perspective. For thermal sources, the LOS from STIX through the maximum of the thermal source is shown~(red~line).}
  \label{fig:STIX_imaging}
\end{figure}

We constructed thermal (6–10~keV) and non-thermal (18–28~keV) STIX images for the two primary peaks in the HXR time profile (numbers 2 and 3) using the mem\_ge algorithm \citep{Massa2020}. X-ray image reconstruction can be challenging. In this case, we were unable to produce robust non-thermal images for the other HXR peaks, which was most likely due to fewer counts in the non-thermal part of the spectrum since the thermal images were not affected. However, we cannot entirely rule out other contributing factors. Figure \ref{fig:STIX_imaging} shows an overview of the STIX light curves with numbered pulsation peaks, the selected time interval for HXR peaks 2 and 3, the corresponding spectra, and the reconstructed thermal and non-thermal images from both the Solar Orbiter/STIX and SDO/AIA perspectives.
    
For HXR peak 2 (first column), both AIA 1600~Å flare ribbons are completely covered by the non-thermal STIX contours (blue). The northern non-thermal STIX source fits the two separate kernels observed in AIA 1600~Å, whose evolution we found to correlate well with the HXR pulsation (see Fig. \ref{fig:time_dist_northern_ribbon}). The southern flare ribbon was found to consist of several UV kernels at this time (Fig.~\ref{fig:time_dist_southern_ribbon}), consistent with the more continuous STIX non-thermal source. The majority of the non-thermal emission appears to come from the western part of the southern ribbon, which we found to be the location of the kernel most correlated with the HXR pulsation.

The LOS of the thermal STIX sources (red line in the bottom two panels) intersects the top of the flare loop arcade in AIA 131~Å and is therefore interpreted as the typical thermal HXR loop top source. An additional small non-thermal source (see the STIX perspective from the first interval) does not coincide with any bright feature in AIA 1600~Å. Its proximity and relative position to the main thermal HXR source may indicate that it is also located in the above the loop top.

During HXR peak 3 (second column), the non-thermal STIX image consists of only a single compact source located at the western end of the southern ribbon, consistent with the compact kernel we observed in UV (Fig.~\ref{fig:time_dist_southern_ribbon}). However, no non-thermal HXR emission from the northern ribbon was detected by STIX, although it appears bright in the included AIA 1600~Å image. Our spatiotemporal analysis (\ref{fig:time_dist_northern_ribbon}), showed that the UV emission at this point is not well correlated with the HXR pulsations and appears quite diffuse and much less bright compared to the compact kernel in the southern ribbon. Therefore, there could have been HXR emission from the northern ribbon at that time, but if it similarly distributed to the UV emission, it was probably not observable with STIX due to its limited dynamic range.

\section{Discussion}\label{sec:discussion}

We analyzed QPPs with a period of about 3.4~min observed during the eruptive M3.7 flare on February 24, 2023 (see Fig.~\ref{fig:event_overview}). In particular, we compared the HXR pulsations observed by Solar Orbiter/STIX with the complex temporal and spatial evolution of the UV pulsations observed by the SDO/AIA 1600~Å channel. In this section we discuss the implications of the main observational results, show how some phenomena are related to the flare geometry, and finally discuss some potential physical mechanisms for the QPPs.

\subsection{Key observational findings}\label{sec:discussion_key_findings}

The UV pulsations in the integrated light curves of selected subregions of the flare appear quite different (see Fig.~\ref{fig:AIA1600_submaps_and_lightcurves}). In general, the UV pulsations are much stronger and better correlated with the HXR pulsations in regions of negative magnetic polarity (southern flare ribbon and eastern filament region) compared to regions of positive magnetic polarity (northern flare ribbon and western filament region).

STIX images (see Fig.~\ref{fig:STIX_imaging}) constructed for the two strongest HXR peaks (2 and 3) show that the positions of the UV kernels in the northern and southern ribbon are consistent with the source positions of the non-thermal HXR emission. The strong spatial correlation between non-thermal HXR and UV emission from the flare ribbon suggests that the UV emission is at least partially a result of heating from the impact of non-thermal electrons, as expected from the standard flare model. Furthermore, the analysis of the STIX spectra during the flare (see Fig.~\ref{fig:STIX_spectral_evolution}) showed that the HXR pulsations are associated with a soft-hard-soft evolution of the electron spectral index, indicating a modulation of the electron acceleration efficiency correlated with the pulsations.

Our time-distance maps of the southern flare ribbon (Fig.~\ref{fig:time_dist_southern_ribbon}, revealed a particularly interesting spatiotemporal evolution of the UV sources. UV pulsations are associated with individual flare kernels, which occur at an expanding front of the flare ribbon. This pulsating ribbon front first moves rapidly westward until HXR peak 4 and then drifts back eastward, forming a continuous path of UV pulsations in the time-distance plots. This shows that these UV pulsations are not stationary and that the QPPs seen in the integrated light curve are actually made up of discrete bursts at different locations, demonstrating the value of such a spatiotemporal analysis. The fact that all UV pulsations occur at an expanding ribbon front shows that they are directly related to instantaneous magnetic reconnection and confirms that this is an apparent motion of the UV kernels driven by the subsequent heating of adjacent parts of the chromosphere as the magnetic reconnection progresses.

During the second HXR pulsation, additional UV kernels brighten and appear to move back east earlier, forming a separate inactive ribbon front (see Figs.~\ref{fig:time_dist_southern_ribbon} and \ref{fig:parameters_southern_ribbon}) without further UV pulsations. This indicates the presence of two magnetic subsystems, of which only the one associated with the more southwestern active ribbon front produced UV pulsations. This implies that only a subset of the total magnetic flux currently undergoing magnetic reconnection was involved in the process causing the pulsations. This further suggests that only the pulsation-active ribbon front was associated with modulation of the electron acceleration efficiency, while the inactive ribbon front was probably associated with a continuous (unmodulated) energy deposition.

Another interesting aspect of the southern ribbon is that the southwestern outline of the final ribbon becomes visible during the first HXR peak (Fig.~\ref{fig:time_dist_southern_ribbon}, top panel and movie). It is further activated after the second HXR pulsation in  a way that ends up producing a faint southward streak in the time-distance plot ($y = 510$ to $500~arcsec$). This early activation thus seems to mark part of the boundary of the total magnetic flux, which was subsequently reconnected during the flare. The early activation of its outline may be related to observations of flare precursors. Together with the inactive ribbon front, this faint southwestern UV emission almost completely outlines the region from which all further pulsations originate at the time of the third HXR peak (see movie). The active ribbon front moves across this outlined region during HXR pulsations 4 to 7 and eventually catches up with the inactive front (see also the color-coded evolution in Fig.~\ref{fig:parameters_southern_ribbon}). This behavior further emphasizes that only a subset of the reconnecting flux is pulsating, and how it is separated from the rest of the flux.

In addition to the general apparent motion of all UV kernels following the expanding active ribbon front, each individual UV kernel associated with HXR pulsations 4 through 7 starts off at the northern end of the current ribbon front and then propagates along it to the south with varying intensities and velocities, resulting in further southward streaks in the time-distance plot (Fig.~\ref{fig:time_dist_southern_ribbon}). This may indicate that the mechanism producing the pulsations also propagates along the reconnection region with each pulsation, or it may imply the presence of other reconnection phenomena such as slipping reconnection.

In the eastern filament region we observed strong UV pulsations in the integrated light curve (Fig.~\ref{fig:time_dist_eastern_filament}, black light curve) as well as strong pulsations in the instantaneous area and the derivative of the cumulative area and flux (Fig.~\ref{fig:parameters_eastern_filament}). In this region, these UV pulsations are not emitted by compact flare kernels, but mainly by hot plasma injected from the direction of the main flare arcade into an additional straight filament channel (Fig.~\ref{fig:time_dist_eastern_filament}, pink light curve; see also the accompanying movie). This distributed emission source makes the pulsations most visible in the summed brightness (Fig.~\ref{fig:time_dist_eastern_filament}, panels three and four).
While we cannot exclude the presence of HXR emission in this region due to the limited dynamic range of STIX, it appears that the plasma was heated somewhere closer to the main flaring region and then moved into the straight filament channel, producing strong UV pulsations in this region. The close temporal correlation with the HXR pulsations suggests that the plasma injections are modulated by the same underlying pulsation mechanism as the HXR pulsations. Thus, the heating is likely still driven by the impact of non-thermal electrons, but the UV emission we observe is spatially separated from the heating location. This observation shows that integrated light curves of a flare can mix multiple emission sources (footpoint sources vs. plasma injected into coronal loops), further emphasizing the importance of spatially resolved studies.

\subsection{Eruption-driven 3D flare reconnection in asymmetric magnetic geometry as a major driver}\label{sec:discussion_3D_flare}

The SDO/AIA observations (see Fig.~\ref{fig:SDO_GONG_overview} and accompanying movie) show that the flare and its associated filament eruption were strongly asymmetric. In the east we observe an extended coronal dimming region which corresponds to the anchor point of the freely erupting side of the filament. This region is surrounded by a J-shaped flare ribbon. These are typical features of eruptive flares: the footpoints of flux ropes are regions where the plasma density drops due to the large-scale expansion of the structure, while flare ribbons are associated with regions of strong current densities \citep{Janvier2014,Janvier2016}. The J-shaped ribbon morphology is expected in the presence of flux ropes, as they trace the photospheric imprint of large connectivity change regions, also known as quasi-separatrix layers \citep[QSLs,][]{Demoulin1996JGR}. The QSLs and current ribbon J-shape morphology is also found in the context of simulations \citep{Janvier2013}, as the photospheric signatures trace the imprint of the 3D current volume \citep[see e.g.,][]{Kliem2013}, which is a fundamental element of the standard flare model in 3D.

However, the 3D standard flare model \citep[see e.g., Fig.7 in][]{Janvier2014} needs to be adapted to the unique magnetic field configuration of the flaring region under study: in some cases, reconnection with surrounding field lines leads to the flux rope footpoints drifting away  \citep[e.g.,][]{Aulanier2019, Zemanova2019, Chen2019}, so that the ribbon morphology and flare loops can deviate from the predictions given by the model. In other cases, a more complex set of flare ribbons can be observed, due to parasitic magnetic regions appearing in the vicinity \citep{Janvier2023}.  Here, the western part of the filament is overlaid by a hot loop system connected to the strong magnetic field region near the sunspots. In the usual 3D standard flare model, the flare loops connect to the straight portion of the J-shaped flare ribbons. Here, they connect to a compact southern flare ribbon and a more elongated northern ribbon, from which they eventually begin to drift westward around a bend in the inverse-S-shaped PIL. We conclude that the overall evolution of the filament structure and ribbon morphology can therefore be simply explained by the 3D standard flare model, when taking into account the specific asymmetric magnetic geometry (bent PIL and overlying loops constraining parts of the erupting filament) of the region.

We now propose a thorough description of the different flare consequences, starting with flare loops.
The initial rapid westward expansion of the southern UV flare ribbon, especially during the second HXR peak, was accompanied by an antiparallel eastward motion of the northern ribbon and a strong-to-weak shear transition of the flare loops observed in SDO/AIA EUV channels. Both antiparallel footpoint motions and flare loop shear transitions are commonly observed features of solar flares and appear to be a natural result of eruption-driven magnetic reconnection in a 3D geometry \citep{Aulanier2012}. 

After this initial shear transition, flare ribbons typically show a more gradual expansion, mostly perpendicular away from the PIL. The southeastward motion we observed is likely caused by a bend in the inverse-S-shaped PIL (see HMI LOS magnetogram in Fig.~\ref{fig:AIA1600_submaps_and_lightcurves}). A drift of the flare loops around this bend is clearly visible at the time of the fourth HXR peak (Fig.~\ref{fig:SDO_GONG_overview}, fourth AIA 131~Å image). Relative to this new orientation, the more gradual southeastward expansion of the southern ribbon during HXR peaks 4 to 7 can be viewed as the second part of the usual ribbon evolution.

This bend could also explain the weak or absent UV pulsations in the positive polarity, namely the northern ribbon and its extension into the western filament region, during the later HXR peaks.
Field lines involved in magnetic reconnection fan out over a wider ribbon area on the northwest side of the bend compared to the compact southern ribbon. The impact region of the non-thermal electrons would then be much more spread out, resulting in less heating and a weaker UV signal, easily obscured by the other activity in that region.

We speculate that the bend might even have caused a slight separation of two flux systems. One connects the southern ribbon to most of the northern ribbon. The other connects the southern ribbon to the more westerly regions. This could explain the observations of two ribbon fronts, with the pulsations present only in the more westerly flux system.

\subsection{Possible pulsation mechanisms}\label{sec:discussion_pulsation_models}

In typical flare observations it can be challenging to unambiguously identify a specific pulsation driver. As summarized in the review by \citet{Zimovets2021}, models often lack testable predictions that clearly distinguish them from others, and available datasets often simply lack the necessary information due to observational limitations including saturation and low cadence \citep[e.g.,][]{Inglis2023}. However, this dataset offered a unique observational testbed since the timescale of the pulsations was sufficiently long to be captured by EUV/UV imagers, e.g., AIA with a cadence of 12/24~s in EUV and UV passbands respectively. In addition, the flaring region did not suffer from significant saturation effects, likely due to its large spatial extent. As a result, we were able to track the spatial origin of the UV and HXR pulsations.

Our key observational findings, outlined in section \ref{sec:discussion_key_findings}, indicate that reconnection and subsequent particle acceleration occurs at various locations along the arcade on the pulsation timescale, in a very characteristic way. We hypothesize that the pulsation timescale and spatial attributes are therefore a signature of the 3D magnetic field structure undergoing reconnection in the flaring region, for the reasons outlined in the following.

The 3D nature of magnetic reconnection in solar flares deviates from the standard “cutting and pasting” of magnetic field lines of the 2D picture. In 3D, QSLs are shown to be regions of drastic changes in the field line connectivity \citep[see][]{Demoulin1996}, generalizing the concept of magnetic reconnection sites from the specific cases of null points and separators. They can be mathematically described by the norm, $N$, obtained by analyzing the divergence of field line connectivity and by calculating the Jacobian matrix associated with the field line mapping \citep{Pariat2012}. QSLs are associated with the preferential locations for strong current densities to arise, especially in regions with the strongest norms \citep{Aulanier2005, Aulanier2012}. There, magnetic field lines are seen to “flip” \citep{Priest1992} or “slip” \citep{Demoulin1996}, due to their rapid but continuous change of magnetic connectivity.

This aspect was investigated in a 3D numerical setup, where a correlation between the slipping reconnection motions and speeds of reconnecting field lines during a flux rope eruption was found with the morphology of QSLs and intensity of the norm \citep{Janvier2013}. In their study, the authors found that the slipping motion speed was a linear function of the mapping norm, with the most drastic changes in the magnetic connectivity (and therefore the highest mapping norm) related to super-Alfvénic speeds, also defined as slip-running reconnection, while the reverse is simply defined as slipping reconnection. The numerical model predictions of slipping motion of field lines were since confirmed by several observations at high SDO/AIA cadence, such as in \cite{Dudik2014}, where the authors showed that the motions of kernels along flare ribbons, as well as the apparent motion of flare loops, could be attributed to the resulting consequences of 3D reconnection in the presence of QSLs. In particular, the motion of kernels in relation with the magnetic field connectivity was investigated in details in \citet{Lorincik2019}. In recent years, high-cadence observations have enabled the characterization of super-Alfvénic connectivity change by investigating the slip-running motion of kernels in flare ribbons \cite{Lorincik2025}. 

Locations where the norm of the connectivity is small can be associated with slower apparent motion of field lines and kernels when reconnection takes place. Since the deposition of energy in the chromosphere is observed along the flare ribbons, this means that the slower the change of connectivity is, the more the energy output related to particles accelerated along the reconnected field lines will be constrained to a specific area in the chromosphere. In contrast, regions of very high connectivity gradients undergoing slip-running reconnection will lead to a fast energy deposition along the ribbons, resulting in a lower average energy flux over a larger surface covered during the same timescale. Therefore, this could explain the discrete HXR sources observed along flare and current ribbons \citep[e.g.,][]{Musset2015}.

Here, our observations consist of near continuous faint ribbon emission combined with periodic localized pockets of bright emission. This spatiotemporal ribbon evolution fits with the signature of 3D reconnection, where the periodicity is directly linked with the characteristics of the reconnection region. The spatial distribution of emission along the ribbons likely reflects the structure of the reconnecting coronal magnetic field and its intersection with the chromosphere, while the areas producing pulsations in the active ribbon can be a signature of slipping reconnection (similarly to the way that kernels are well-defined spatial structures). Thus, both are seen to result from the nonsmooth connectivity change of the magnetic field. It is evident that this would then lead to inhomogeneity in the ribbon brightening distribution. The spatial evolution of the chromospheric imprint would therefore depend on how the coronal magnetic structures of the flaring region evolves over the course of the flare which is linked to the filament/CME dynamics in the case of eruptive flares.

While the 3D nature of the magnetic reconnection explains the observed spatial evolution of the energy deposition sites and can also explain localized enhancements of this energy deposition, it is unclear whether it can account for the quasi-periodic nature of this modulation and how regular enhancements can occur as the ribbons and associated QSLs move away from the PIL (pulsations 4 to 7). For this reason, it may still be necessary that some additional feedback mechanism or oscillation was active in tandem with the evolving 3D reconnection. Indeed, while the slipping reconnection mechanism is directly related to connectivity mapping of magnetic field lines, we also need to consider the physics of the reconnection within the 3D current density volume. Here, the presence of turbulence, or inhomogeneous current density, could also play a role in the dynamics of energy release and its imprints in the different layers of the Sun's atmosphere. Furthermore, the reconnection process itself within the current layer can have an oscillatory behavior. Several models have been proposed for how eruption-driven magnetic reconnection could develop pulsations. For example, the termination shock could develop instabilities \citep{Takahashi2017} or the current sheet itself could develop a series of plasmoids due to the plasmoid instability \citep[e.g.,][]{Loureiro2007,Bhattacharjee2009}. However, direct observational evidence for these models (e.g., plasmoid observations) will necessitate further developments in terms of our capability to resolve reconnection sites, which remain inaccessible with the current instrumentation.

In principle, all mechanisms that either modulate the efficiency of the energy release process via MHD oscillations (group ii in \citet{Zimovets2021}) or enable a spontaneous quasi-periodic energy release (group iii) appear to be viable options. In contrast, mechanisms that directly modulate the emission (group i) are inconsistent with or observations. Any pulsation mechanism involving an oscillating stationary flare loop (e.g., standing sausage, kink, or slow mode) is incompatible with the UV kernel motions. Furthermore, the observed soft-hard-soft evolution of the HXR spectrum is better explained by a direct modulation of the energy release.

If the energy release process is modified by such an additional mechanism, the apparent path of the pulsations would still be driven by the propagation of magnetic reconnection, including slipping reconnection, as discussed above. However, the quasi-periodic enhancement of the reconnection process by an additional driver would periodically increase the deposited energy at the expanding ribbon front. This periodic enhancement, superimposed on the motion of the ribbon, could also lead to the observed spatiotemporal evolution of the UV kernels. However, identifying the exact mechanism that could have been active during this event is beyond the scope of this study and probably unfeasible from the available data set.

Given the 3.4~min pulsation period identified for this event, it would be tempting to link these pulsations to 3 min oscillations observed in sunspots (see e.g., the review by \citet{Bogdan2006}). In the literature, pulsations of this timescale have often been attributed to a modulation of the energy release by the waves emitted from these sunspot oscillations \citep[e.g.,][]{Sych2009,Kumar2016,Li2024}. However, the extent and location of the flare ribbons relative to the sunspot make it difficult to conceptualize a direct relationship between sunspot oscillations and the characteristics of the ribbon kernels, as they are directly related to the energy deposition due to the flare occurring well above the sunspots. In such a scenario, it might also be difficult to explain how it is only a subsystem of the reconnection that was modulated.

\subsection{Implication and future direction}\label{sec:discussion_implications}

We have shown how UV pulsations can have a complex spatial and temporal evolution, how they can vary widely between different subsystems of the flare, and how different emission processes can be active at the same time. Our results show how pulsations need to be considered as part of the 3D reconnection geometry and emphasize the importance of spatially resolved studies of QPPs.

Of course, the feasibility of such studies is limited by the cadence and resolution capabilities of (E)UV imagers. In this case, the pulsation period of about 3.4~min for this flare was quite long compared to the 24~s imaging cadence of the AIA 1600~Å channel. Combined with the large spatial extent of the event, this has enabled much of the analysis we performed for this study. It also highlights the utility of unsaturated, high-cadence observations of flare ribbons at UV and EUV wavelengths, as well as of X-ray imagers with improved dynamic range, for  studies of flare energy release. Similar studies could be performed for pulsations with shorter periods using IRIS \citep{DePontieu2014_IRIS} and high-cadence, short-exposure images of EUI/HRI \citep{Rochus2020_EUI} gathered during recent dedicated flare campaigns \citep{Ryan2025}. The utility of unsaturated short-exposure EUI/FSI images for identifying footpoints and relating them to HXR emission sources has already been demonstrated by \citet{Collier2024b}. These findings further highlight the immense potential of these EUI/HRI imaging campaigns for future pulsation studies.

\section{Conclusion}\label{sec:conclusion}

We analyzed UV and HXR pulsations with a period of about 3.4~min observed during the eruptive M3.7 solar flare of February 24, 2023. Our results reveal the significant spatiotemporal complexity of the UV pulsations, as demonstrated by these key findings:

\begin{itemize}
    \item The strength of UV pulsations and their correlation with HXR pulsations varied greatly between different flare subregions (main flare ribbons and filament anchor points in both polarities).
    \item The most prominent and spatiotemporally complex UV pulsations occurred in the compact southern flare ribbon. They were emitted by a series of flare kernels that occurred at an expanding flare ribbon front. Later kernels also propagated individually along the ribbon front, adding a further fine structure to the observed pulsation pattern.
    \item A strong spatial correlation between these UV kernels and HXR sources suggests that non-thermal electrons were the primary driver of both emissions. The HXR spectrum exhibited a soft-hard-soft evolution on the same 3.4~min timescale, indicating a modulation of electron acceleration efficiency with each pulsation.
    \item In contrast, another expanding front of the southern ribbon showed no pulsations, implying that only a subset of the reconnecting flux was involved in the mechanism producing pulsations.
    \item UV pulsations near the eastern filament anchor point were emitted from plasma injected into a separate filament channel. While likely driven by the same underlying mechanism, these pulsations were emitted in a spatially separate way with respect to their heating source, in contrast to the flare kernels.
\end{itemize}

The overall flare geometry strongly reflects expected key features of a 3D reconnection process in the presence of QSLs, while much of the asymmetry and complexity in our observations seems to be the result of an asymmetric magnetic confinement (e.g., bend in the PIL, overlying loops structure above the western leg of the filament). We conclude that much of the difference in the behavior of UV pulsations across the flare has also been the result of this specific magnetic structure.

The detailed spatiotemporal behavior of UV pulsations in the southern ribbon seem to be closely linked to the progression of magnetic reconnection, including slipping reconnection, in this asymmetric confinement. We further hypothesize that the geometry of the associated QSL and its fine structure in combination with slipping reconnection could have contributed significantly to the formation and motion of the localized UV brightenings. To explain the full observations, we hypothesize that this mechanism was active in combination with a time varying reconnection process.

Our results highlight the spatiotemporal complexity of QPPs and the need to consider them as part of the 3D magnetic structure of the flare. This demonstrates the importance of spatially resolved studies to separate different emission sources and fully capture the dynamics of QPPs. Only such detailed observations allow us to follow the UV pulsations on the energy release time scales and explore their connection to the evolving reconnection process. This study was enabled by the relatively long pulsation period (3.4~min) and the large spatial scale of the event, allowing for the detailed tracking of UV sources. Future investigations could find it useful to focus on shorter period QPPs, using high-cadence observations from instruments such as IRIS and EUI/HRI, to further explore the role of 3D reconnection processes in driving QPPs.

\begin{acknowledgements}
Solar Orbiter is a space mission of international collaboration between ESA and NASA, operated by ESA.
The STIX instrument is an international collaboration between Switzerland, Poland, France, Czech Republic, Germany, Austria, Ireland, and Italy. 
This research was funded in part by the Austrian Science Fund (FWF) 10.55776/I4555. For the purpose of open access, the author has applied a CC BY public copyright license to any Author Accepted Manuscript version arising from this submission. L.A.H is funded through a Royal Society-Research Ireland University Research Fellowship. The research was sponsored by the DynaSun project and has thus received funding under the Horizon Europe programme of the European Union under grant agreement (no. 101131534). Views and opinions expressed are however those of the author(s) only and do not necessarily reflect those of the European Union and therefore the European Union cannot be held responsible for them.
\end{acknowledgements}

\bibliographystyle{aa}
\bibliography{bib-file}

\begin{appendix}

\section{Mask evolution with higher threshold}\label{sec:appendix_masks_and_parameters}

\begin{figure}[ht]
  \resizebox{\hsize}{!}{\includegraphics{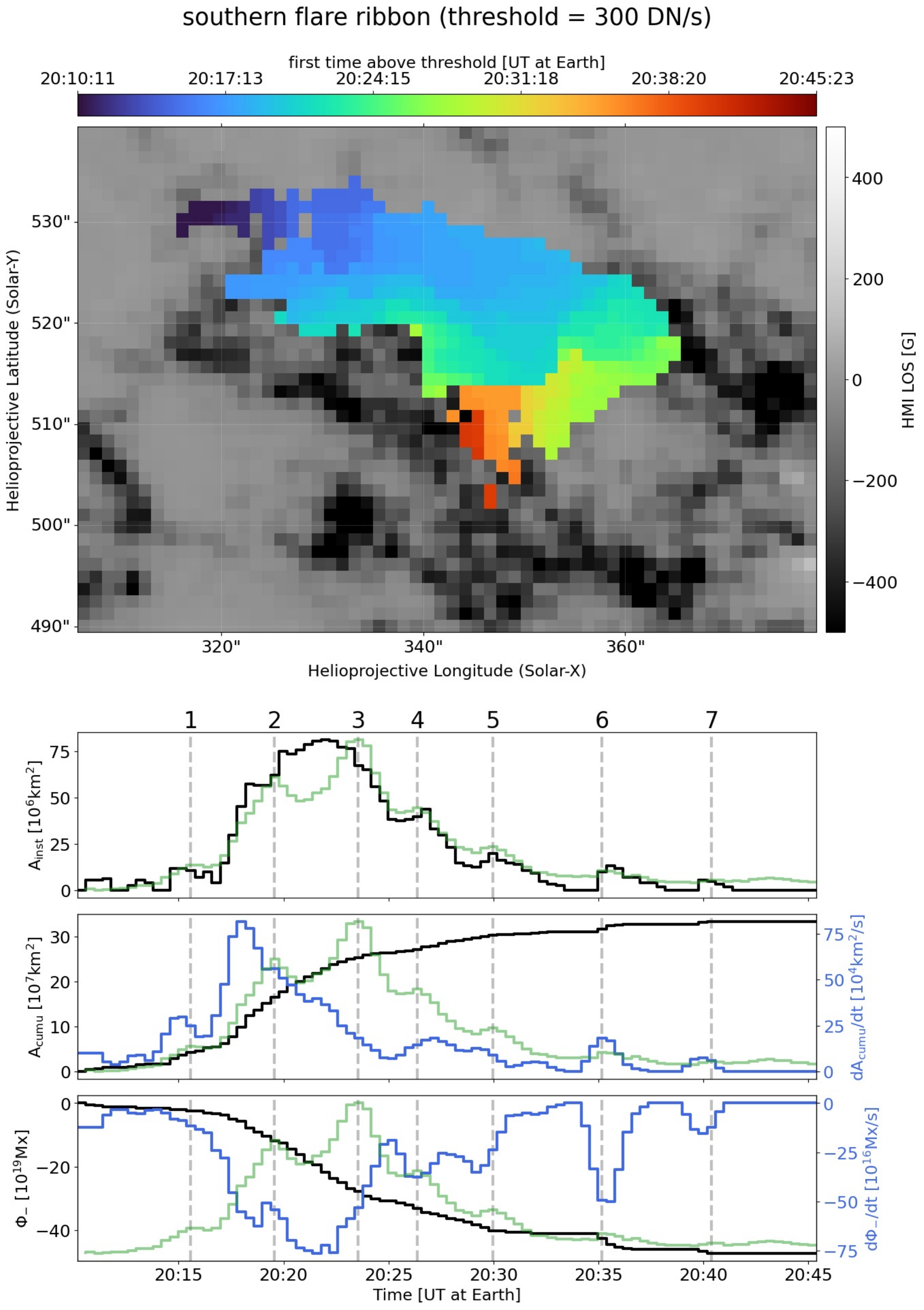}}
  \caption{Same as Fig. \ref{fig:parameters_southern_ribbon}, but with a higher threshold of 300 DN/s.}
  \label{fig:parameters_southern_ribbon_higher_threshold}
\end{figure}

Figure~\ref{fig:parameters_southern_ribbon_higher_threshold} complements the results of the mask evolution and extracted parameters for the southern ribbon shown in Section \ref{sec:results_AIA1600_masks_and_parameters}. In this version, we applied a higher threshold of 300~DN/s, which highlights the increase in correlation between the extracted parameters and the HXR pulsations with increased threshold. 

\end{appendix}

\end{document}